\documentclass[sigconf,nonacm]{acmart}
\AtBeginDocument{%
  }

\newcommand{\added}[2][]{#2}
\newcommand{\deleted}[2][]{}
\newcommand{\replaced}[3][]{#2}

\usepackage{graphicx}
\usepackage{booktabs}
\usepackage{subcaption}
\usepackage{array}        

\usepackage{xcolor}
\usepackage{framed}
\usepackage{fancyvrb}

\definecolor{shadecolor}{gray}{0.95} 
\begin{document}


\title{Interview-Informed Generative Agents for Product Discovery: A Validation Study}



\author{Zichao Wang}
\affiliation{%
  \institution{Adobe Research}
  \city{San Jose}
  \state{CA}
  \country{USA}}
\email{jackwa@adobe.com}

\author{Alexa Siu}
\affiliation{%
  \institution{Adobe Research}
  \city{San Jose}
  \state{CA}
  \country{USA}}
\email{asiu@adobe.com}


\newcommand{\alexa}[1]{\textcolor{purple}{[alexa: #1]}}
\newcommand{\jw}[1]{\textcolor{blue}{[jw: #1]}}




\begin{abstract}
Large language models (LLMs) have shown strong performance on standardized social science instruments, but their value for product discovery remains unclear. We investigate whether interview-informed generative agents can simulate user responses in concept testing scenarios. Using in-depth workflow interviews with knowledge workers, we created personalized agents and compared their evaluations of novel AI concepts against the same participants’ responses. Our results show that agents are distribution-calibrated but identity-imprecise: they fail to replicate the specific individual they are grounded in, yet approximate population-level response distributions \deleted[id=jw2]{after re-matching}. These findings highlight both the potential and the limits of LLM simulation in design research. While unsuitable as a substitute for individual-level insights, simulation may provide value for early-stage concept screening and iteration, where distributional accuracy suffices. We discuss implications for integrating simulation responsibly into product development workflows.
\end{abstract}




\begin{teaserfigure}
  \centering
  \includegraphics[width=0.75\textwidth]{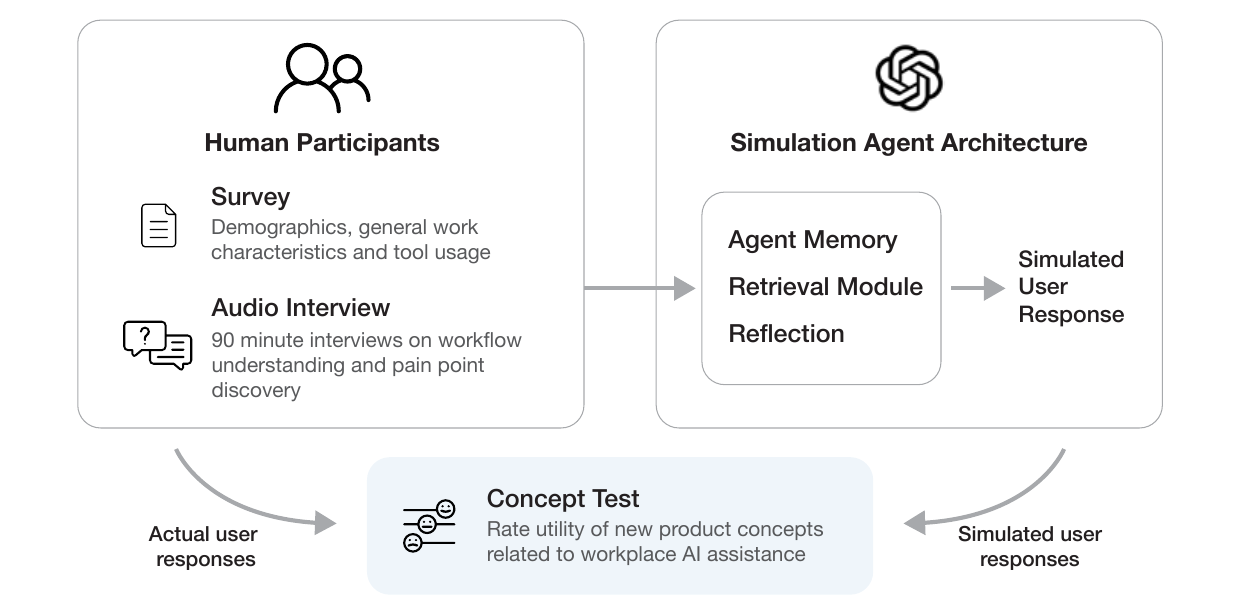}
  \caption{Study design and simulation setup. We recruit knowledge workers to conduct in-depth interviews and surveys. This data is used to ground our generative simulation agents. To evaluate the agents' performance, we compare their responses to those of human participants in a concept testing session. This allows us to explore the application of these agents in product discovery contexts.}
  \Description{Enjoying the baseball game from the third-base
  seats. Ichiro Suzuki preparing to bat.}
  \label{fig:teaser}
\end{teaserfigure}

\maketitle

\section{Introduction}

Large language models (LLMs) have demonstrated capabilities on \deleted[id=jw2]{certain} standardized social science instruments, \added[id=jw1]{with performance varying considerably across instruments and studies,} raising questions about their potential for human behavioral simulation in research contexts~\cite{MurraySmith2022,2208.10264,2209.06899}. Some recent work has shown promising accuracy on established measures like survey responses and personality assessments, suggesting possible applications for user research~\cite{1000,2402.01765,huang-etal-2024-reliability}. This presents a potential opportunity for product development and design discovery tasks, where understanding user responses to new concepts is valuable but traditional user studies, while rigorous and essential, require significant time and resources. In this work, we investigate the applicability of LLM simulation specifically in product discovery contexts. 

\replaced[id=jw3]{Park et al.~\cite{1000} recently showed that interview-informed agents can reproduce individuals’ responses to social-science instruments with high normalized accuracy, e.g. achieving 85\% normalized accuracy on General Social Survey items relative to humans’ own two-week test–retest consistency. Our work builds on this architecture but probes a different application setting: early-stage product discovery. While product discovery shares some characteristics with social science survey simulation, it also presents presents an application context that has received less empirical validation for LLM simulations than the social science contexts. Rather than predicting attitudes on long-standing constructs such as ideology, personality, and fairness, we ask whether interview-informed agents can simulate how knowledge workers respond to {\it novel, hypothetical AI workflows}, including both scalar ratings such as Technology Acceptance Model (TAM)~\cite{Davis1989} and Net Promoter Score (NPS)~\cite{Reichheld2003} as well as open-ended design feedback. In this setting, preferences are being {\it constructed on the fly} for artifacts participants have never used, and designers care about both population-level patterns (which concept is promising?) and identity-level heterogeneity (which kinds of users are excited or blocked?). While both social-science and product-discovery settings mix established scales with bespoke questions, the latter setting places different demands on simulations: they must generate actionable, contextualized rationales about new designs~\cite{Griffin1993,10.5555/2821566,Page1992}, not only reproduce responses on standardized instruments. }{Park et al~\cite{1000} found that interview-informed generative agents could replicate participant responses on General Social Survey items with 85\% accuracy when grounded in detailed life-history interviews. A key aspect of their approach was leveraging rich qualitative interviews to inform their simulation agents. We adopt this methodology but seek to validate its utility in a product discovery context.
Their evaluation also focused on established social science instruments with standardized questions and response formats. While product discovery shares some characteristics with social science survey simulation, it also presents presents an application context that has received less empirical validation for LLM simulations than the social science contexts. Rather than measuring established constructs with known ground truth, product concept testing requires generative, exploratory responses about novel artifacts. Users must evaluate workflow fit, identify unmet needs, and provide actionable design insights—tasks that demand contextual reasoning about their specific work practices and pain points~\cite{Griffin1993,10.5555/2821566,Page1992}.}

\added[id=jw1]{Concretely, we investigate two research questions about this product discovery setting:}
\begin{itemize}
    \item \textbf{\added[id=jw1]{RQ1 (Scalar fidelity).}} \added[id=jw1]{To what extent can interview-informed agents reproduce human scalar ratings at both the individual level and the population level in product concept tests?}
    \item \textbf{\added[id=jw1]{RQ2 (Qualitative alignment).}} \added[id=jw1]{In what ways do agents’ open-ended responses resemble or diverge from participants’ own qualitative feedback?}
\end{itemize}

\added[id=jw1]{Section~\ref{sec:results} provide preliminary empirical evidence to answer the above questions. Finally, we draw on these findings to discuss implications (Section~\ref{sec:discussion}) and limitations (Section~\ref{sec:limitation}) for when such simulations may usefully augment (rather than replace) traditional product discovery practice.}

We conduct a focused validation study to examine whether interview-informed generative agents can effectively simulate user responses to product concepts, bridging the gap between demonstrated LLM capabilities and practical product development needs. Figure~\ref{fig:teaser} shows an overview of our work.
We study this, through a case study of AI document workflow tools. We conducted workflow-focused interviews with knowledge workers ($N=51$), then created personalized agents and compared their concept evaluations to the same participants' actual responses.
We assessed simulation performance using quantitative measures (\replaced[id=jw3]{TAM, NPS}{Technology Acceptance Model, Net Promoter Score}) and qualitative response analysis across four AI concept prototypes. \added[id=jw1]{In total, we collected 3,060 responses (4 concept prototypes, 15 question each, 51 users). The generative agents simulated the same number of these responses, which forms the basis of our analysis and findings.}

Our findings indicate that agents achieve population-level distributional similarity while showing limited individual-level accuracy. Interview-informed agents approximate human response distributions and outperform baseline approaches, yet fail to reliably simulate specific individuals they are designed to represent. \deleted[id=jw2]{When re-matched to better-fitting participants, agents show improved performance, suggesting they may capture general user types rather than specific individuals~\cite{Tjuatja2024,jiang-etal-2025-language}}
These results have implications for both practice and research. For practitioners, the findings suggest potential utility for early-stage concept screening, where population-level trends may suffice for initial design decisions. Teams could leverage simulations to rapidly explore multiple design directions and reduce their focus to promising concepts, while still relying on authentic user interviews for detailed insights and validation. For researchers, these findings highlight the importance of domain-specific validation and the need to establish appropriate fidelity thresholds for different research objectives~\cite{Wang2025,Kapania2025}.

\added[id=jw1]{
We summarize our contributions as follows.}
\begin{itemize}

    \item \textbf{\added[id=jw1]{Empirical validation of interview-informed agents for product discovery.}}
    \added[id=jw1]{We provide, to our knowledge, the first systematic evaluation of interview-informed generative agents on early-stage product concept testing, combining TAM, NPS, and open-ended feedback for four AI document workflow concepts.}

    \item \textbf{\added[id=jw1]{Characterization of simulation fidelity in this setting.}}
    \added[id=jw1]{We show that agents are \emph{distribution-calibrated but identity-imprecise}: they approximate human response distributions and outperform baselines at a population level, yet fail to reliably match the specific individuals they are designed to represent.}

    \item \textbf{\added[id=jw1]{Implications for product discovery practice.}}
    \added[id=jw1]{We discuss practical guidance on when such simulations may be useful (e.g., low-cost concept screening and directional trade-off exploration) and when authentic user interviews remain essential (e.g., understanding individual workflows, trust, and adoption barriers).}

\end{itemize}

\section{Related Work}

\subsection{Generative AI for User Simulation}

Recent research explores agent-based simulations for replicating human behavior across a range of contexts. Applications include usability testing, design feedback, and public opinion polling~\cite{Xiang2024,Lu2025,Ataei2025,2501.15283}; generating synthetic personas and role-playing diverse user groups for scalable studies~\cite{genagents,1000,2503.16527}; rapid prototyping and ideation in design~\cite{10.1145/3392850}; and large-scale simulations for policy, economics, and social science~\cite{2412.03563,anthis2025position,Kapania2025}. These efforts highlight both the appeal and the challenges of using generative agents in research. Most relevant to our work, Park et al.~\cite{1000} show that interview-informed agents can approximate human survey responses. We extend this line of inquiry to the domain of product discovery, testing whether interview-grounded agents can simulate participants in concept evaluation tasks. To our knowledge, this is the first validation of generative agents in this setting.

The design of our simulation agents builds on a recurring “memory–retrieval–reflection” architecture~\cite{2210.03629,genagents,1000,2502.07443,Lu2025}, with reasoning supported by either the underlying foundation model~\cite{2501.12948} or chain-of-thought prompting~\cite{2201.11903}. Prior studies differ in how much information they use to instantiate agents: some rely on minimal demographic attributes~\cite{2406.20094}, while others show that richer inputs, such as hours-long interviews\added[id=jw1]{~\cite{1000}} \added[id=jw1]{and foundation model fine-tuning techniques~\cite{Binz2025,suh-etal-2025-language}}, yield better fidelity\deleted[id=jw2]{~\cite{1000}}. Our approach follows the latter, investigating whether detailed interview transcripts improve simulation quality in product discovery.

Despite their promise, generative agents face well-documented limitations. They can misrepresent or essentialize demographic groups, raising risks of epistemic injustice~\cite{Wang2025}; fail to respect participant agency and consent~\cite{2508.06950}; or mis-simulate basic causal reasoning~\cite{Gui2023}. Other studies find that simulated behaviors diverge from authentic human patterns~\cite{2503.16527,2501.15283}, produce overly detailed or unrealistic outputs~\cite{Lu2025}, and overrepresent Western cultural norms and professions~\cite{2501.08579}. Our work contributes to this discussion by evaluating both the limitations and the potential of interview-informed simulations in the context of product discovery.

\subsection{Product Discovery and Concept Evaluation}
\added[id=jw1]{
In product discovery, teams ask potential users to evaluate novel or not-yet-built concepts based on sketches, scenarios, or simple prototypes rather than fully functional systems. Decades of work in consumer behavior and judgment and decision making treat these early evaluations as forecasting tasks under uncertainty, where people form expectations about future benefits, effort, and risk using heuristics and prior experience~\cite{Tversky1974,Kahneman1979,Oliver1980}. HCI and design research has operationalized this through methods such as contextual inquiry and scenario-based design~\cite{Carroll2000}, experience prototyping~\cite{buchenau2000experience}, and technology or cultural probes~\cite{hutchinson2003technology,gaver1999cultural}, which encourage participants to imagine how a speculative system would fit into their everyday practices. Early concept evaluation studies further show that even when only rough representations are available, people can meaningfully judge overall “goodness” and high-level qualities such as pragmatic and hedonic experience, and that these judgments are relatively robust across different representation formats (e.g., text, pictures, video, or simple prototypes)~\cite{10.1145/1868914.1868991,Eidloth2023}.}

\added[id=jw1]{Technology adoption frameworks such as the Technology Acceptance Model (TAM) and Diffusion of Innovations conceptualize these early judgments in terms of perceived usefulness, ease of use, and compatibility with existing practices, which are robust predictors of behavioral intention and eventual adoption~\cite{Davis1989,Rogers2003}. Complementary satisfaction frameworks such as expectation–disconfirmation and Net Promoter Score (NPS) capture downstream loyalty and recommendation intent~\cite{Oliver1980,Reichheld2003}. Value Sensitive Design and related approaches further emphasize probing stakeholders’ values, possible harms, and envisioned use contexts through mock-ups, prototypes, and field deployments before a technology is fully realized~\cite{friedman1996value}. Taken together, this literature frames product discovery as eliciting open-ended, contextual, and counterfactual reasoning about how a novel concept might integrate into real workflows and how people trade off benefits and burdens. This complements the LLM simulation literature in social science, which has focused on predicting responses to standardized, established instruments (e.g., attitudinal scales, economic games) where the task is to select from fixed response options about existing behaviors. Our study builds on this product discovery literature by asking whether interview-informed generative agents can approximate this more anticipatory form of reasoning in the context of early-stage AI document workflow concepts, complementing prior work that has primarily evaluated simulations on standardized survey instruments.
}

\section{Methodology Overview}

Our approach uses in-depth user interviews to create LLM-based user simulations for product concept testing. We first conduct detailed interviews with target users to capture their workflows, pain points, and technology adoption patterns. These interviews inform personalized generative agents that simulate how each user would evaluate novel product concepts. We validate simulation accuracy by comparing agent responses to actual user responses from concept testing sessions, using both quantitative measures (TAM, NPS) and qualitative feedback analysis. This methodology enables rapid exploration of multiple concepts while maintaining grounding in authentic user perspectives. Our case study demonstrates the approach using AI document workflow tools evaluated by knowledge workers. In Section~\ref{sec:case_study_design}, we discuss the case study design and interview protocols. In Section~\ref{sec:simulation_setup}, we discuss how we leverage the study data for user simulation.

\section{Case Study Design}
\label{sec:case_study_design}

{\bf Domain: AI document workflows}. \added[id=jw1]{Our choice of the domain is inspired by recent work showing that document-heavy workflows represent a key bottleneck for knowledge workers with potential for AI-powered document tools to streamline productivity~\cite{fok2024marco, yun2025generative, han2022passages, jahanbakhsh2022understanding}, where opportunities to design better and more suitable human-AI collaborative tools, affordances, and workflows abound.} As a case study, we focus on knowledge workers who spend substantial time reading and analyzing complex documents (e.g., contracts, financial reports, research papers, business proposals, regulatory documents, etc). We focused on roles in Finance, Legal, Operations, Management, and Research functions where document analysis is central to job performance. \deleted[id=jw2]{Recent work has shown that document-heavy workflows represent a key bottleneck for knowledge workers with potential for AI-powered document tools to streamline productivity~\cite{fok2024marco, yun2025generative, han2022passages, jahanbakhsh2022understanding}.}

\begin{figure*}
    \centering
    \includegraphics[width=0.8\linewidth]{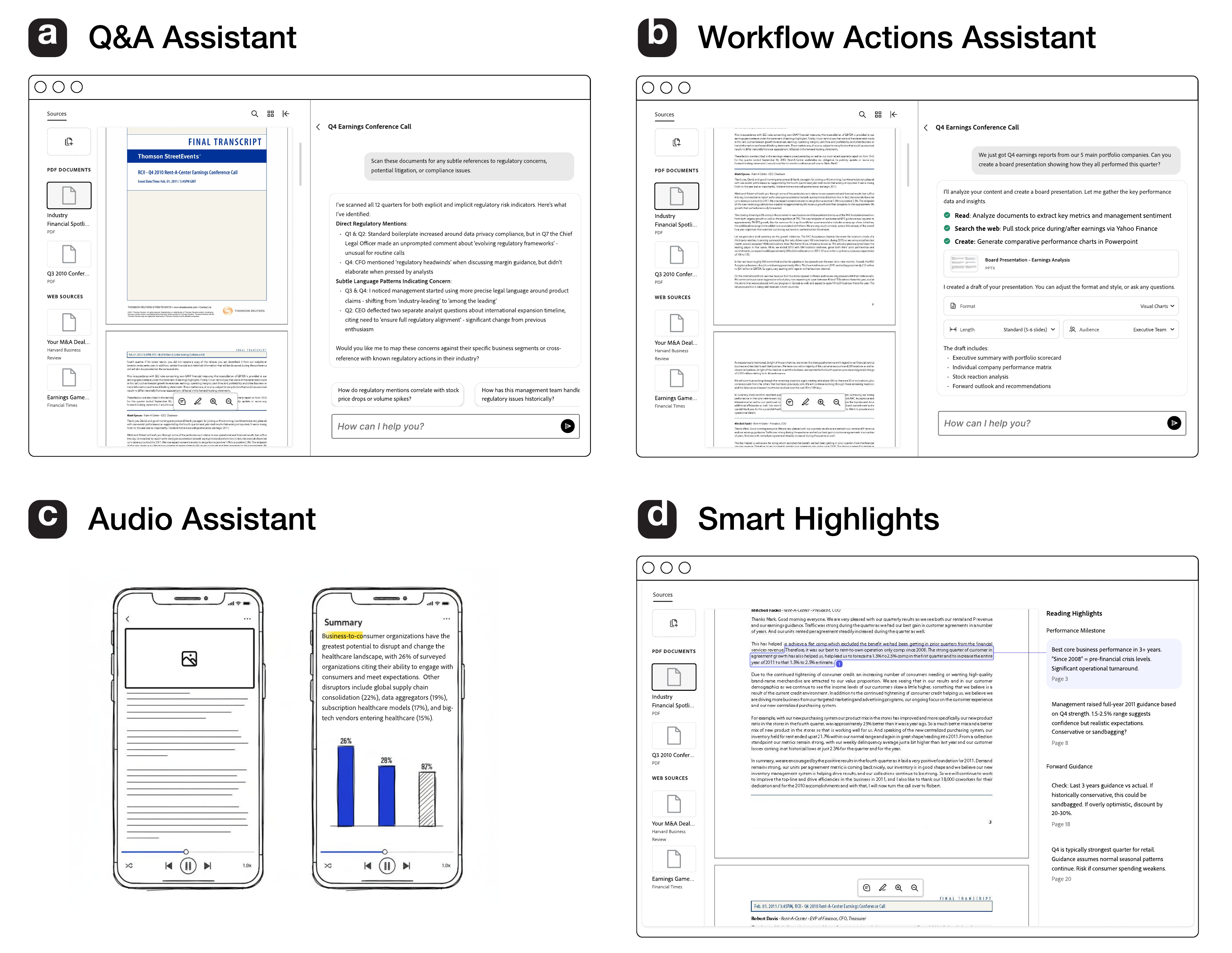}
    \caption{Prototype concepts for AI document workflow assistance. The four designs span different levels of intervention and integration: (a) Multi-document Q\&A Assistant, (b) Smart Highlights Assistant, (c) Audio Assistant, and (d) Workflow Actions Assistant. These concepts represent a spectrum from passive support to active automation, enabling evaluation across varied user needs. \added[id=jw1]{Higher resolution figures are available in the Supplementary Material D.}}
    \Description{description}
    \label{fig:concepts}
\end{figure*}

\textbf{Concept test prototypes}. \added[id=jw1]{Inspired by recent advances in AI document assistants such as those that can transform documents into podcast experiences~\cite{google_notebooklm} and perform question and answering (Q\&A) over document(s)~\cite{chatpdf}, } \replaced[id=jw3]{w}{W}e selected four concepts that span different interaction paradigms and workflow integration levels, designed to test various dimensions of AI assistance in document workflows:
\begin{enumerate}
    \item Multidoc Q\&A Assistant: Analyze multiple sources to answer questions with grounded response (Figure~\ref{fig:concepts}a).
    \item Smart Highlights Assistant: Auto-identify key information with contextual margin notes (Figure~\ref{fig:concepts}b).
    \item Audio Assistant: Transform documents into interactive audio with voice navigation (Figure~\ref{fig:concepts}c).
    \item Workflow Actions Assistant: Detect tasks and execute multi-step actions automatically (Figure~\ref{fig:concepts}d).
\end{enumerate}

These concepts were chosen to represent a spectrum from passive assistance (highlights) to active automation (workflow execution), and from individual document focus (audio) to multi-document synthesis (Q\&A), allowing us to test how different levels of AI intervention that might resonate with different users.

\subsection{Data Collection Protocol}


We conducted a longitudinal study with knowledge workers to establish the effectiveness of interview-informed simulation for product concept testing. The study comprised four phases designed to capture rich user context and later enable validation of simulation accuracy.

\subsubsection{Study Design Overview}
\mbox{}\par
\textbf{Screening survey}. An initial online questionnaire identified participants with document-heavy workflows suitable for evaluating AI document assistance concepts. Participants were screened for at least weekly PDF usage, specific document types that required long-form reading (contracts, financial reports, research papers, business proposals, regulatory documents), and relevant job functions (Finance, Legal, Operations, Management, Research). 

\textbf{Part 1: in-depth interview} (90 minutes). 
Interviews were conducted as unmoderated audio sessions deployed through Qualtrics, allowing participants to complete them at their convenience while ensuring consistent protocol delivery. Interviews explored participants' current document workflows, technology adoption patterns, pain points, and AI tool experiences. 
The interview protocol was designed to capture the contextual information necessary for creating realistic user simulations, including detailed use case walkthroughs, workflow mapping, and perceptions of AI capabilities. \added[id=jw1]{On average, each participants contributed about 43 minutes of raw audio recording time, and each response on average contains 106 words. In total, we collected over 3,000 concept-test responses and about 36 hours of speech from these participants.}

\textbf{Part 2: initial concept testing} (30 minutes). 
Participants evaluated four AI concept prototypes using standardized measures and open-ended feedback through Qualtrics. Responses were audio-recorded. This session was completed within 24 hours of the interview to maintain contextual continuity while avoiding user fatigue from an overly long single session. Concept Testing responses serve as the ground truth for validating simulation accuracy.

\textbf{Part 3: follow-up concept testing} (30 minutes). 
Participants re-evaluated the same concepts using identical procedures as Part 2. This session occurred 3 days after Part 2 to assess participants' self-consistency in their evaluations, following established protocols for measuring individual response reliability in behavioral simulation research~\cite{1000}. 




\subsubsection{Participant Recruitment}
Recruitment was conducted through Prolific\footnote{https://www.prolific.com} with compensation of 25 USD per hour across all study phases. Payments are made to participants only after participants completed all three parts of the study, which we make clear in the recruitment description. We believe that this payment structure incentivizes participants to carefully consider whether to contribute to our study or not and to fully commit if they decide to contribute. In total, 51 participants were recruited using targeted screening criteria and after manual review by the authors. All participants were based in the United States, were knowledge workers, and worked with documents at least daily. Our sample included professionals across diverse job functions, with the largest groups in Management/Leadership (14/51, 27\%), Operations/Administration (8/51, 16\%), and Education (4/51, 8\%). Other represented roles included Research \& Development, Sales, Human Resources, Marketing, Finance, Information Technology, Healthcare, and Legal functions. Participants ranged in age from 21 to 76 years (M = 43.2, SD = 11.9), with an even gender distribution (\replaced[id=jw3]{26/51}{26/61}, 51\% male and 25/51, 49\% female). Regarding technology adoption patterns, participants were predominantly cautious adopters who wait to see how others use new tools before trying them (23/51, 45\%) or early adopters who try new tools quickly (21/51, 41\%). Most participants held positive attitudes toward AI tools for document workflows, with 43\% (22/51) reporting very positive perceptions and 35\% (18/51) somewhat positive, while only 18\% (9/51) were neutral and 4\%(2/51) somewhat negative.


\subsubsection{Concept Test Measures}

We collected both quantitative and qualitative data to enable systematic validation of simulation accuracy while capturing the nuanced insights essential to product concept evaluation. Our measurement approach combined established quantitative instruments with open-ended responses. 

For quantitative assessment, we employed the Technology Acceptance Model (TAM)\replaced[id=jw3]{~\cite{Davis1989}}{~\cite{thomas1974methodology}}, using an abbreviated 6-item scale that measures three core constructs: Perceived Usefulness, Perceived Ease of Use, and Behavioral Intention. Each construct was assessed using two items on 7-point Likert scales (1 = Strongly Disagree, 7 = Strongly Agree), with construct scores calculated as the mean of constituent items following standard TAM scoring procedures. Additionally, we included the standard Net Promoter Score (NPS) question as \replaced[id=jw3]{a common}{an industry-standard} metric of user advocacy~\cite{Reichheld2003}. Together, TAM and NPS provide complementary perspectives on technology acceptance and user satisfaction that are directly relevant to product concept evaluation.

For each concept, participants provided open-ended responses addressing workflow fit, specific use cases, implementation concerns, and improvement suggestions. We also collected comparative evaluations across concepts, asking participants to articulate relative preferences and positioning rationales. 

\begin{figure*}
    \centering
    \includegraphics[width=0.8\linewidth]{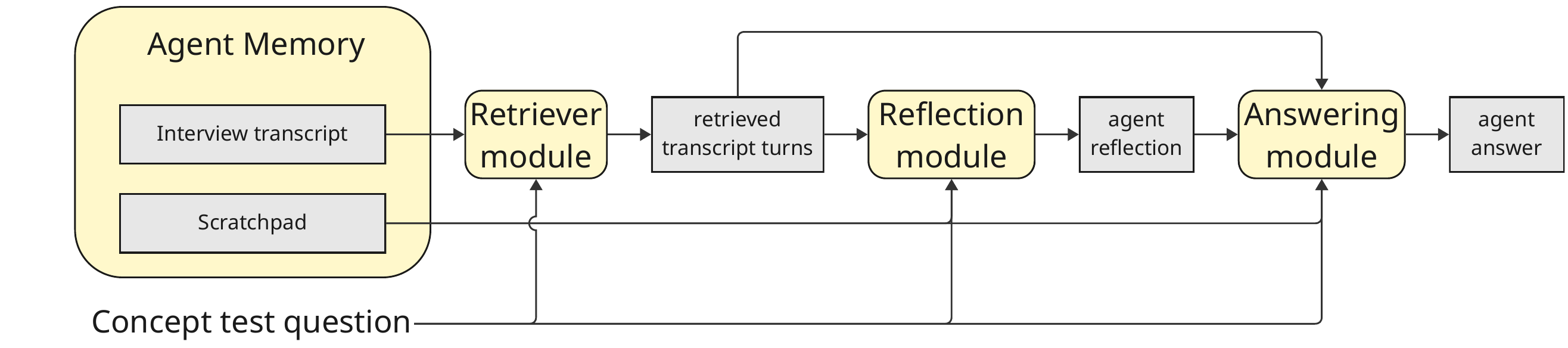}
    \caption{Simulation architecture for interview-informed generative agents. The agent creation module builds memory and scratchpad representations from interview transcripts, while the agent interaction module retrieves relevant turns, reflects, and generates answers to concept test questions. Together, these components define how each agent simulates a target participant.}
    \label{fig:sim-arch}
\end{figure*}

\section{Simulation Methodology}
\label{sec:simulation_setup}

Our simulation approach is as follows. 
For each participant, we use their interview transcript to create their agent. Then, with each agent, we conduct the same concept test ``interview'' that the human participants already did (Figure~\ref{fig:teaser}). We then evaluate the performance of the agents by comparing their answers with their corresponding human participants' answers. The subsections below delve\deleted[id=jw2]{s} into the agent architecture design and evaluation methodology. 

\subsection{Simulation Architecture}

We largely adopt the methodology in~\cite{genagents,1000} to design the agent architecture. Figure~\ref{fig:sim-arch} illustrates the overall architecture, which is organized into four modules: agent memory, retriever, reflection, and answering. Together, these modules define how the agent is constructed from interview data and how it produces responses during simulation.

{\bf Agent memory module.}
The agent memory stores the grounding data that defines each simulated participant. When the interview transcript is available, \replaced[id=jw3]{w}{W}e segment it into conversational turn pairs, each consisting of an interviewer question and a participant response. Each turn pair is then embedded into a vector representation using a text embedding model, and the resulting pairs and their corresponding embeddings form a searchable memory database. In addition, we encode basic demographic information collected in the screening survey (such as job division and title) into a scratchpad represented as a dictionary. The memory database and scratchpad together capture both the detailed interview context and high-level participant attributes.

{\bf Retriever module.}
When posed a new question, the agent first consults its memory via a retriever model. After embedding the new question into a vector, the retriever searches (we simply do maximum inner product search) the vectorized database for the most relevant turn pairs to the current query. These retrieved memories, along with the scratchpad, provide the contextual evidence on which the agent bases its response. This retrieval step ensures that answers remain grounded in the specific transcript of the participant being simulated while avoiding sending the entire memory database.

{\bf Reflection module.}
Given a question and the retrieved memory, the agent generates a reflection. Reflection prompts the agent to synthesize its understanding of the simulated participant: drawing on the scratchpad and relevant memory, the agent articulates reasoning or perspective beyond a direct answer. Reflections may be added back into the memory, becoming retrievable in subsequent interactions. This mechanism allows the agent to gradually build a more coherent internal model of the participant over the course of a session.

{\bf Answering module.}
Finally, the agent produces an explicit answer to the question based on the retrieved memory, the scratchpad, and the generated reflection. For categorical prompts (e.g., Likert-scale or NPS items), the agent outputs a scalar value on the given scale. For open-ended prompts, it generates a free-form textual response.

For all experiments, we first create the agent memory using OpenAI’s text-embedding-3-small model.\footnote{\url{https://openai.com/index/new-embedding-models-and-api-updates/}} Then, when conducting simulated concept test interview, we ask the agent to reflect first and then answer\deleted[id=jw2]{s} the question. Both reflection and direct answering components are implemented using GPT-4o~\cite{gpt4o}.

\subsection{Evaluation Methodology}
\subsubsection{Metrics.}
\label{sec:metrics}
Since our participant data corpus included both quantitative and qualitative measures, we extend the evaluation methodology in~\cite{1000} and assess simulation performance across two dimensions: (1) quantitative alignment between the simulation agent and participant responses, and (2) qualitative similarity in the insights and reasoning patterns extracted from their respective open-ended answers.

For question items that require categorical responses, including those on a 7-point Likert scale and NPS, we assess how closely the agents replicate their corresponding human participants on an individual level using both the mean absolute error ({\bf MAE}) and correlation ({\bf corr}), adopted from~\cite{1000}. \replaced[id=jw3]{We also use Gwet’s AC2, a chance-corrected ordinal agreement coefficient widely used for measuring reliability on Likert data, to compute individual-level response agreement. AC2 quantifies the extent to which two response vectors assign the same ordinal categories beyond chance. We choose this metric because some participants do not use all 7 response categories (or 11 in NPS), variance-based correlations (e.g. Pearson~\cite{1000}) become undefined.}{For correlation computation, we choose Gwet's AC1 as the metric to improve numerical stability where both the human participants and their corresponding agent selects the same answer category across all question items under inspection (other popular metrics such as Pearson correlation~\cite{1000}, Cohen's Kappa~\cite{McHugh_2012_kappa}, and Krippendorff's Alpha~\cite{Krippendorff2011Alpha} are all undefined under the above special circumstances).} \added[id=jw1]{We additionally compute Spearman's correlation suitable for ordinal data by mixing and normalizing the Likert-scale and NPS-scale responses to the same scale between 0 and 1, similarly to how correlation is computed in~\cite{1000}. Finally, we compute the accuracy of exact response match and compute the percentage that different agents achieve relative to human-human agreement.} In the main results, we compute the metric per participant using all their responses and then aggregate over participants to obtain the mean, 95\% confidence interval, and the p-values via bootstrapping. 

We also assess how closely the agents replicate their corresponding human participants on a population level using Wasserstein distance ({\bf WD})~\cite{Kantorovich1960}, a metric that measures the distance of probability distributions appropriate for ordinal responses. Specifically, we collect the responses from all participants (or agents) and compute the distance between their response distributions.
For all the above metrics, we report additional detailed analysis in the Supplementary Materials that further breaks down the statistics and visualizations into finer-grained levels such as at the prototype concept and construct levels.

For question items that require open-ended responses, we compare human participants' and their agents' responses along four qualitative dimensions, including sentiment direction, explanation alignment, topic coverage, and voice and tone similarity. Specifically, we present the evaluator the question along with the two responses from human participant and agent, respectively, and ask the evaluator to output a score along a 7-point Likert scale as well as an explanation. These scores are averaged across the participants to get a mean and confidence interval which we use to report aggregated results. In practice, we adopt LLM-as-a-judge for this qualitative evaluation, using both GPT-4o and GPT-4.1-mini as the evaluators. The two judges achieve a Cohen's Kappa of 0.53, suggesting moderate agreement. For each question item, we average the scores by these two judges before aggregation. 
Detailed prompts and additional results for these evaluations are available in the Supplementary Materials.

\subsubsection{Baselines.}
{\sloppy
In the main paper, we primarily compare three agent designs, including an agent with both interview transcript and scratchpad ({\bf interview-based}), an agent with only scratchpad ({\bf scratchpad-only}), and an agent without transcript nor scratchpad ({\bf no-information}). Our primarily objective is to investigate whether having interview transcripts improves the simulation performance. Each of these agent choices are compared to their human participants' responses to obtain results according to the aforementioned metrics. We also experimented with a range of hyper-parameter choices for the interview-based agent itself, including the number of memories to retrieve and the number of reflection items, along many other choices. We report these ablation study results in the Supplementary Materials. In addition, we also compute all the above metrics among the participants themselves based on the two times they answered the same concept test questions during our longitudinal study. This human-human comparison results establish the human consistency that we use to benchmark the performances of our agent designs.
}

\section{Experiments and Main Results}
\label{sec:results}

\begin{figure*}
    \centering
    \includegraphics[width=0.3\linewidth]{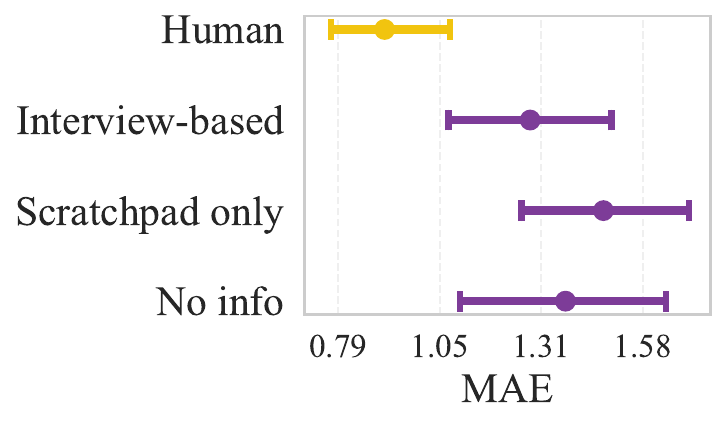}\hspace{20pt}\includegraphics[width=0.3\linewidth]{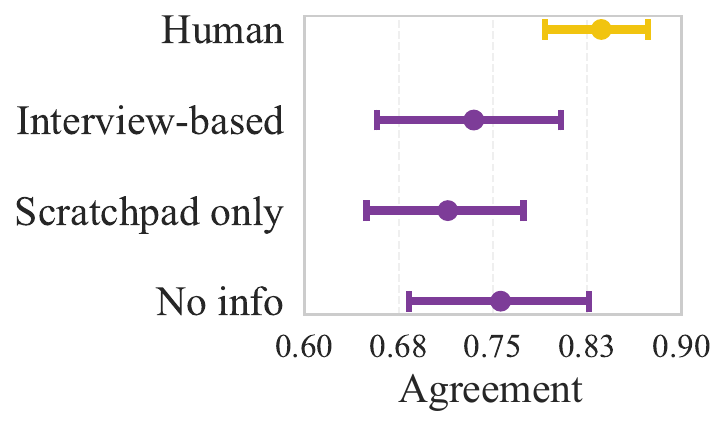}\hspace{20pt}\includegraphics[width=0.3\linewidth]{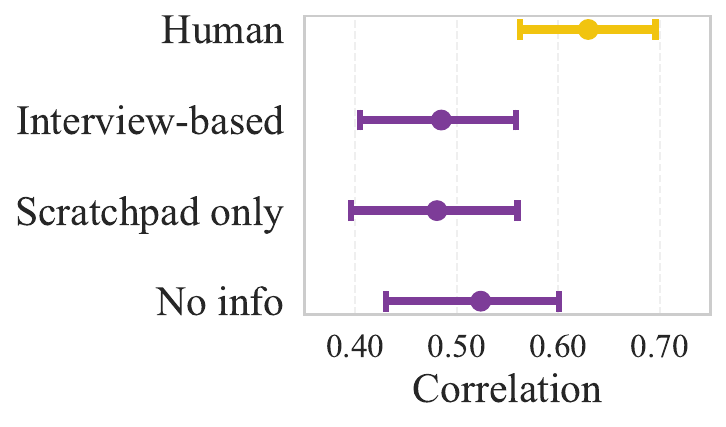}
    \caption{Individual-level accuracy of simulated responses to categorical questions. Compared to human participants, all agent variants show higher error (MAE) and lower \added[id=jw1]{agreement/}correlation, with no significant differences across designs. \added[id=jw1]{Error bars indicate 95\% bootstrap confidence intervals over participants (more details in Section~\ref{sec:metrics}).} \deleted[id=jw2]{Results suggest limited fidelity in replicating specific individuals, even when grounded in full interview transcripts.}}
    \label{fig:main_mae_corr}
\end{figure*}

{\bf Interview-based agents are imprecise on an individual level.} Figure~\ref{fig:main_mae_corr} shows the performance of different agent designs compared to the performance of human participants through both MAE\added[id=jw1]{, agreement,} and correlation metrics. We observe that all agents have significant gap with human participants' performance, and that agents' performances comparing with each other is not significant, as confirmed with a Tukey's post-hoc test~\cite{Tukey1949}. \added[id=jw1]{The accuracy results (numbers in parenthesis show the percentage relative to human-human performance), showing human vs. human, human vs. interview-based agent, human vs. scratchpad only agent, and human vs. no information agent, are 0.446, 0.300 (67\%), 0.259 (58\%), and 0.256 (57\%), implying the best agent design (interview-based) is able to match about 67\% of human performance.} These results suggest that the agents are imprecise on an individual level, unable to accurately replicate their corresponding human participants' answers to categorical questions, even with rich information such as the full interview transcript that we collected from the human participants. Interestingly, we observe that even the agent without any information from the human participant (i.e. no interview transcripts nor scratchpad) is comparable to agents with some information; its average correlation is even slightly higher than the interview-based agent. 

\begin{figure*}
    \centering
    \includegraphics[width=0.8\linewidth]{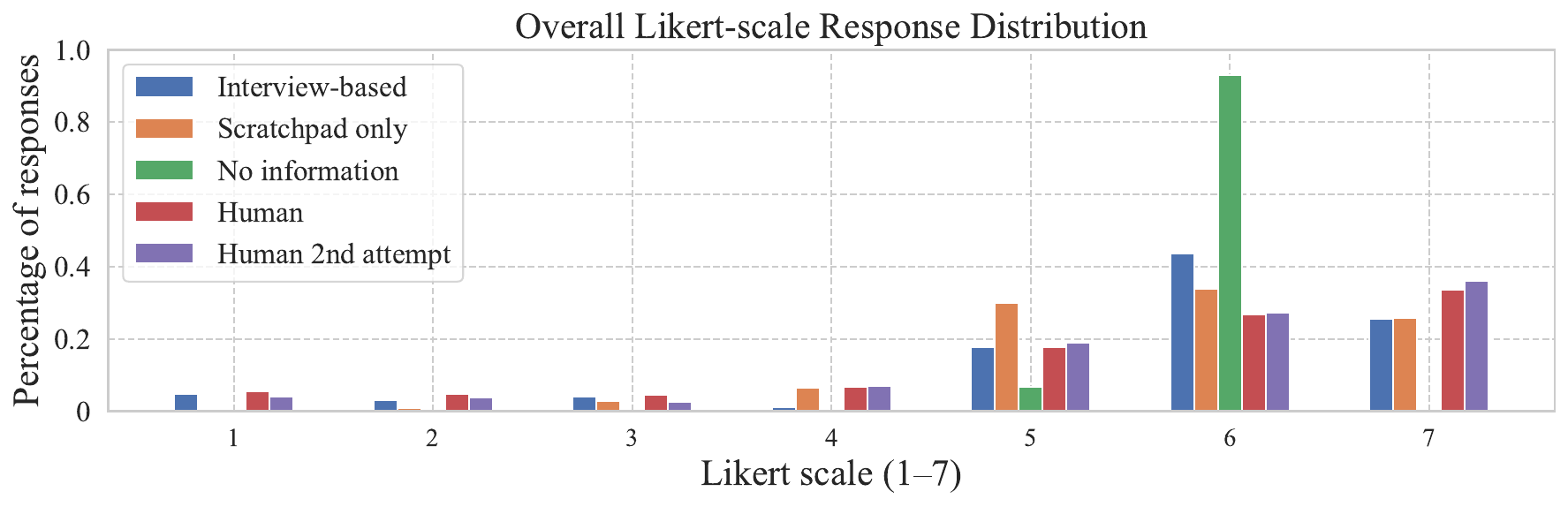}
    \vspace{5pt}
    \includegraphics[width=0.8\linewidth]{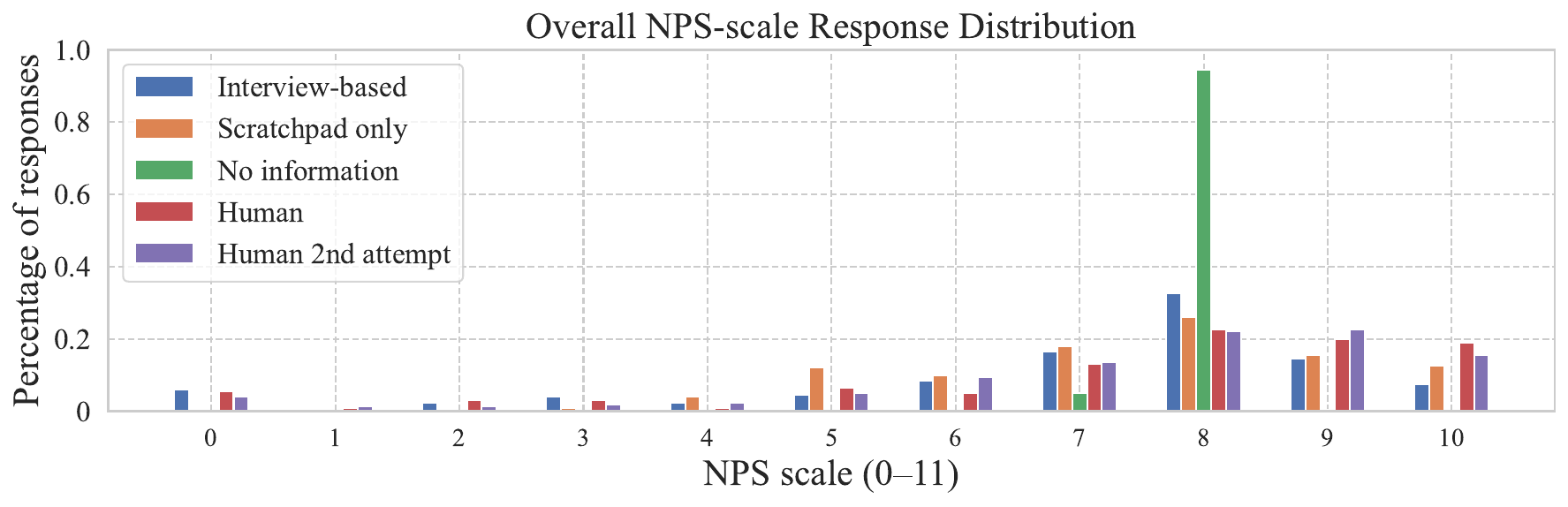}
    \vspace{10pt}
    \begin{tabular}{lcc}
\toprule
& {Likert} & {NPS} \\
\midrule
Human -- Human 2nd attempt & 0.175 & 0.211 \\
Human -- Interview-based & 0.227 & 0.487 \\
Human -- Scratchpad only & 0.393 & 0.513 \\
Human -- No information & 1.058 & 1.678 \\
\bottomrule
\end{tabular}
\vspace{-10pt}
    \caption{\added[id=jw1]{Overall population-level alignment of simulated responses. The two bar charts of categorical responses, Likert-scale and NPS-scale respectively, show interview-informed agents capture more realistic variability compared to scratchpad-only and no-information baselines. The bottom Summary statistics table using Wasserstein distance
    confirm that interview-informed agents achieve the best distributional similarity to both human response attempts compared to alternative agent designs.}}
    \label{fig:combo1}
\end{figure*}
{\bf Interview-based agent is reasonably well-calibrated on \replaced[id=jw3]{a}{an} population level.} 
Although individual-level accuracy remains poor, interview-based agents achieve better population-level alignment than alternative designs.
Figure~\ref{fig:combo1} shows the response distributions as bar charts \added[id=jw1]{(for Likert- and NPS-scale responses respectively)} for human participants across both concept testing sessions and all three agent \replaced[id=jw3]{types. We}{types.We} observe that the agent without any information degenerates.
Detailed analysis in the Supplementary Materials reveals that the no-information agent collapses to almost always choose 6 \replaced[id=jw3]{on}{for} 7-point \replaced[id=jw3]{Likert-scale}{Likert scale} questions and 8 \replaced[id=jw3]{on the 11-point NPS question}{for 11-point NPS question}.
Scratchpad-only agent already performs much \replaced[id=jw3]{better, as}{better as} its curve is more aligned with that of the human participants.
\replaced[id=jw3]{Interview-based agents have the best alignment, especially for negative responses.}{Interview-based agent has the best alignment.} \replaced[id=jw3]{For example, they are the only design that captures lower-scoring preferences, as seen in the close match of the bars between 1 and 2 on the $x$-axis for the Likert plot and 0 and 3 on the x-axis for the NPS plot.}{In particular, it is the only agent design that captures lower scoring preferences as can be observed in the close match of the curves between 0 and 4 on the x-axis.}
\replaced[id=jw3]{The bottom table in Figure~\ref{fig:combo1} provides quantitative validation, showing that Interview-based agent achieves the best similarity compared to human responses in terms of the Wasserstein distance 
compared to alternative agent designs.}{Table~\ref{tab:main-dist-stat} provides quantitative validation using Wasserstein distance, a metric appropriate for ordinal responses.
Tukey post-hoc comparisons show interview-based agents as the only variant with statistically equivalent distributions to human responses (non-significant p-values in both comparisons).}\added[id=jw1]{ This pattern is consistent with findings in other domains where models can be well-calibrated at the population level yet have limited utility for individual-level prediction, and where group-level associations do not straightforwardly translate to individuals~\cite{VanCalster2019Calibration,Baldwin2021ACEs,Freedman1999Ecological}. \added[id=jw1]{We also present the same plots and statistics for each construct and each concept tested in Figures~\ref{fig:bigblock1} --~\ref{fig:bigblock4} in the Supplementary Material.} }

\begin{figure*}
    \centering
    \includegraphics[width=\linewidth]{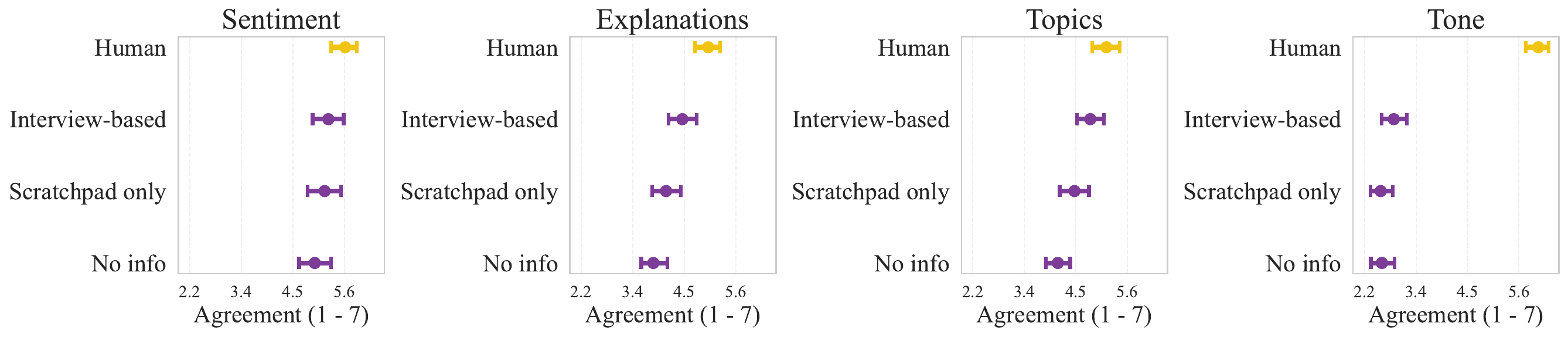}
    \caption{Evaluation of open-ended response quality. Across sentiment, explanation, topic coverage, and tone, interview-informed agents outperform other agent variants but remain consistently below human participants. The largest gap arises in tone, reflecting agents’ inability to capture conversational style from transcripts. 
    }
    \label{fig:main-llm-eval}
\end{figure*}

{\bf Open-ended responses reveal further gaps in response quality between agents and human participants.} Figure~\ref{fig:main-llm-eval} shows participant-level aggregated qualitative results for each of the four metrics. We can observe that the interview-based agent performs better than the remaining agent designs across metrics but significantly underperforms human participants (confirmed by Tukey's post-hoc test). Of the four metrics, the agents perform the worst on the voice and tone metric, suggesting that none of the agent designs adequately captures how the human participants are speaking. This \replaced[id=jw3]{could}{cound} be due to the different formats of the data recorded: data from human participants are transcribed from speech whereas data from agents are generated as if the interviewee is writing the response rather than speaking. We also show additional results in the Supplementary Materials comparing a pair of example responses from a human participant and from their corresponding interview-based agent, along with the LLM judge's reasoning and score on the voice and tone metric, to illustrate the difference in the tone and feel of the responses. 

\added[id=jw1]{We also conduct thematic analysis comparing human responses and agent responses. Here we use the comparison between human and the interview-based agent in the context of ``Audio Assistant'' concept as a case study for illustration. Our findings reveal that human participants and AI ``participants'' produced strikingly similar headline themes but notably different texture. Across both sets of responses, we saw the same core opportunity areas: using audio to handle long, dense PDFs while commuting or multitasking (``reviewing contracts while commuting'' or when eyes are tired from long research sessions), accuracy and privacy as the dominant barriers (especially for technical, medical, or compliance content), and a strong desire for integration with existing tools like Office, EHRs, or project-management systems. In this sense, the simulated agents are distribution-calibrated: a designer reading only the AI simulated responses would still recover the same top-level themes about when audio helps and what must be in place (accuracy, security, integration) for adoption. However, the human responses add layers of nuance that the AI rarely reproduces. Approximately 30\% of human participants expressed outright disinterest or rejection (``I would not use this... in any circumstances at work''; ``I specifically would not use this situation at all'' and multiple other similar expressions), while only a single AI-simulated participant expressed comparable level of outright skepticism (``I don't really see many situations where audio would be preferable to reading documents in my current workflow''). Humans articulated sharp misfits with their actual work practices, such as the mobile-only format being a dealbreaker for desktop-based workers (``I'm not gonna be able to do this on my computer, which is where I'm gonna be doing most of my work. I'm not gonna be doing that from a mobile phone'') and personal cognitive preferences that fundamentally misaligned with the concept (``I'm not a big audiobook guy... I need to read the actual book myself''). They also described what ``accuracy'' means in practice, such as mispronounced medical terms, wrong ratios in financial reports, reputational risk, and raised organizational constraints (admin approval, compliance policies, cost justifications) that the AI responses tended to flatten into generic statements about ``strong security measures.'' Human participants also demonstrated authentic confusion about the concept's mechanics (``I'm a little confused on how that works''), whereas AI agents consistently produced clear, structured interpretations. Perhaps most tellingly, AI responses exhibited uniformly constructive framing: even negative feedback was couched in language suggesting potential adoption given improvements. In contrast, humans produced responses ranging from enthusiastic endorsement to dismissive rejection to philosophical tangents (``I'm so sick of seeing people on their cell phones all the time... you don't know if they're on Instagram... I feel more inclined to do business with people who are actually on their computers.''). Overall, we characterize the simulated responses as achieving high-level thematic but not experiential fidelity: they recover the same axes of benefit and risk at a population level, but they lack nuances and smooth over the contradictions, emotions, and messy workarounds that make human interviews useful for understanding real-world adoption and non-adoption.}

{\bf \added[id=jw1]{Cost analysis.}} \added[id=jw1]{To illustrate the pragmatic value of utilizing agents, we compare the time and cost to conduct a concept test between a human participant and the interview-based agent. For human participants, each concept test takes 30 minutes and costs 12.5 USD. For the agent, each concept takes about 4 minutes to run and costs 1.27 USD. For the latter, pricing is estimated from the official OpenAI API documentation}\footnote{\small\added[id=jw1]{\url{https://platform.openai.com/docs/pricing}}}\added[id=jw1]{ and GPT4o which we use for the majority of our experiments. We believe with the advances in foundation models, the time and cost of simulation will further be reduced.}

\section{Discussion}
\label{sec:discussion}

Our simulation results reveal significant limitations: interview-based agents fail to achieve adequate fidelity with their corresponding human participants and show no significant advantage over simpler agent architecture designs. 
\replaced[id=jw3]{Compared to prior work on social-science-style survey simulation such as Park et al.~\cite{1000}, which reported higher individual-level alignment on well-established instruments under a carefully validated interview protocol, our agents perform notably worse in this product discovery setting. Rather than treating this as a definitive contradiction, we interpret it as early evidence that simulation fidelity is sensitive to domain, measurement, and implementation details, including the nature of the task (product concepts vs.\ social attitudes), the specific instruments (TAM, NPS, and concept-specific items), and the interview protocol, rather than as a categorical difference between “social science” and “product discovery.” Our sample size (51 participants across four concepts) places limits on the precision of our estimates and on the strength of our conclusions, and therefore we treat these results as initial evidence about this simulation pipeline that should be replicated at larger scale and with additional concepts.
}{
These findings contrast with established results in social science simulation, where interview-informed agents demonstrated promise on standardized instruments like the General Social Survey~\cite{1000}. This divergence demonstrates a critical insight—simulation techniques that succeed in one domain may not transfer to others, particularly when moving from well-structured survey contexts to exploratory product discovery scenarios.}

\added[id=jw1]{One simple explanation consistent with our data is that the agents exhibit a largely monotonic bias on the rating scales (e.g., systematically under- or over-estimating scores) rather than arbitrary noise. Our current analysis intentionally compares raw scores without post-hoc calibration so as to establish a conservative lower bound on fidelity and to reflect how practitioners might initially use such simulations. Future work should explore normalization and mathematical transforms (e.g., learned linear or nonlinear mappings from agent scores to human scores) to compensate for such bias and test whether these adjustments substantially improve agreement.
}

The performance gap that we observe between human participants and agents also highlights the importance to further advancing the simulation techniques and agent design to close the gap. Recent advances in large language and vision-language models, agent architectures, information retrieval methodologies, and agentic memory designs could all contribute to improved simulation performance; we leave such explorations as an exciting direction for future work. Our results on no-information agent aligns with results in existing literature, suggesting that the agent tends to degenerate and collapse to the ``population mean'' when little or no information is provided~\cite{Wang2025, 2508.06950}. 

Should we stop using simulations for product discovery tasks due to the poor individual-level fidelity? Our population-level results suggest that there may be a path forward with the current interview-based agent design: as long as we only care about the marginal distribution of the categorical responses, which still reveals important insights on how people react to certain concepts on a population rather than individual level, then the population of agents could still be useful in this context. We show that interview-based agent is the only agent design that achieve comparable distribution with human participants, suggesting that interview transcripts are still important to achieve high-quality simulation. The implication is that carefully and painstakingly collecting human data such as interview transcript might be a necessary and unavoidable step; shortcuts such as only collecting demographic information may degrade the simulation performance.

\paragraph{\added[id=jw1]{When simulations may be useful.}} \added[id=jw1]{In light of these strengths and limitations, we see a constrained but potentially valuable role for interview-informed simulations in product discovery. Because agents approximate population-level response distributions but not individual responses, they are better aligned with early-stage questions such as ``Which of these concepts appears more promising overall?'' than with fine-grained personalization. For example, teams might use simulations to run low-cost concept screening and directional trade-off exploration across many variants (e.g., different automation levels or framings of the same workflow) before investing in more expensive prototyping and recruitment. In this role, agents are a complement to, rather than a replacement for, traditional user research, providing fast, approximate signals about relative concept appeal under carefully controlled prompts.}

\paragraph{\added[id=jw1]{When authentic user interviews remain essential.}} \added[id=jw1]{The same findings also clarify when simulations should not be used as a substitute. The poor individual-level fidelity, together with the limitations we observe in qualitative alignment, suggest that interview-informed agents are ill-suited for understanding specific participants’ workflows, trust dynamics, and adoption barriers. Detailed questions such as ``Why does this participant hesitate to delegate this task?'' or ``How would this tool fit into their existing document ecosystem?'' require rich, situated accounts that our agents do not reliably reproduce. For these purposes, authentic interviews and observational studies remain necessary, with simulations serving at most as a way to generate hypotheses or candidate questions rather than definitive evidence.}

\added[id=jw1]{Finally, our findings should be interpreted in light of the study’s scope. We focus on 51 knowledge workers, four AI document-workflow concepts, and a specific interview-informed agent architecture. As such, our results are best viewed as a case study of generative agents in early-stage concept testing for AI productivity tools, rather than a general statement about all product categories or user populations. We hypothesize that sample size, domain stakes (e.g., healthcare vs. productivity), and concept complexity may all shape simulation performance, and we identify varying these factors as an important direction for future work.}

\section{Limitations}
\label{sec:limitation}
Our study has several limitations that point toward important directions for future research.

{\it Sample \replaced[id=jw3]{S}{s}ize and \replaced[id=jw3]{S}{s}cope.} We conducted 51 interviews, which is adequate for a product discovery case study but small relative to large-scale social science simulations~\cite{Xiao2020,1000}. This limits statistical power and the diversity of participant profiles we could cover. Additionally, while we implemented screening criteria, we acknowledge that participants' self-report of document-heavy workflows cannot be independently verified, and this remains a limitation of online recruitment. Future work should examine larger and more heterogeneous populations, as well as multiple product domains beyond document workflows, to test the generality of our findings.

{\it Agent \replaced[id=jw3]{D}{d}esign \replaced[id=jw3]{C}{c}onstraints.} Our simulation agents used a relatively simple architecture based on retrieval, reflection, and GPT-4o generation. Variations in memory representation, retrieval strategies, reflection mechanisms, and foundation model choices could significantly affect outcomes. Exploring and exploiting latest developments in these areas and conducting systematic ablation studies will help clarify which components matter most for simulation fidelity.

{\it Dependence on \replaced[id=jw3]{I}{i}nterview \replaced[id=jw3]{P}{p}rotocol.} The quality and format of the human interviews used to ground the agents likely shaped the outcomes. Semi-structured interviews may capture workflows and pain points differently than ethnographic methods, diary studies, or ongoing contextual inquiry. Future work should explore how different grounding protocols—including automated or AI-mediated interviews—impact simulation quality, and whether richer modalities (visuals, audio, embodied tasks) can reduce the mismatch we observed.

{\it Evaluation \replaced[id=jw3]{M}{m}etrics.} Although we leverage metrics such as MAE, correlation, Wasserstein distance, as well as LLM-assisted thematic coding and evaluation, these measures capture only part of what matters in design discovery. \added[id=jw1]{In particular, LLM-as-a-judge evaluations are themselves subject to bias and model-specific failure modes, so our results using them should be seen as complementary to, rather than replacements for, human-coded analyses and call for further advances in LLM-based automatic evaluation methodologies.} More nuanced metrics of creativity, contextual grounding, and actionability of insights may better capture where simulations succeed or fail. Future work should also investigate how practitioners actually use agent outputs in real design cycles, rather than evaluating simulations in isolation.

{\it Fidelity \replaced[id=jw3]{V}{v}ersus \replaced[id=jw3]{U}{u}tility.} Our results show that agents are “distribution-calibrated but identity-imprecise”: they fail to replicate the specific participant they were meant to simulate, but can approximate group-level distributions. This raises questions about what level of fidelity is “good enough” for different product research tasks. Future work should investigate when distributional accuracy suffices (e.g., early concept screening) and when individual-level fidelity is indispensable (e.g., persona-specific usability testing).

\section{Conclusion}
Our study reveals that interview-informed agents achieve population-level calibration while failing at individual-level replication: they cannot accurately simulate specific users but instead can approximate group response patterns.
This suggests simulation is unlikely to replace individual-level research, but may augment early-stage product discovery by revealing distributional trends. Practitioners can leverage simulations to rapidly explore multiple design directions and reduce focus to select promising concepts, while relying on authentic user interviews for detailed insights and validation.
For researchers, these findings highlight the importance of domain-specific validation and the need to establish appropriate fidelity thresholds for different research objectives. 

\bibliographystyle{ACM-Reference-Format}
\bibliography{references}

\appendix

\section{Interview and Concept Test Protocol}

\paragraph{Goal} Build a comprehensive understanding of knowledge workers document workflows and perceptions to AI tools. The data collected will be used for data simulation. 
\begin{itemize}
    \item     Understand user habits, preferences, and decision-making processes when working with documents 
    \item     Understand how users measure success and efficiency in their document workflows 
    \item     Explore current usage, perceptions, concerns, and interest in AI tools for document workflows 
\end{itemize}

\paragraph{Recruitment criteria.}
\begin{itemize}
    \item     Works with documents almost daily as part of work 
    \item     Regularly works with at least one of:     Contracts and legal agreements,   Financial reports and statements 
    Research papers and academic articles,
    Business proposals and regulatory documents 
    \item     Job division: Finance, Procurement / Purchasing, Management / Leadership, Legal, Operations / administration, Research / Applied research 
    \item     Balance recruitment across people who use and don’t use AI tools to support their work 
\end{itemize}

\subsection{Screening Survey}
Below we list the original questions that we use to select and recruit human participants in our study.
\begin{itemize}
    \item In what division is your job? Please select one.
    \begin{itemize}
        \item     Marketing  
        \item Finance  
        \item Procurement / Purchasing  
        \item Human resources  
        \item Management / Leadership  
        \item Sales  
        \item Research \& Development 
        \item Legal  
        \item Operations / administration  
        \item Other. Please specify:
        \item None of the above  
    \end{itemize}
    \item Briefly describe your current job title: <free text response> 
    \item Which of the following software tools do you use regularly in your job? (Select all that apply) 
    \begin{itemize}
        \item     Microsoft Office Suite 
        \item Google Workspace 
        \item Slack/Teams 
        \item CRM software 
        \item Project management tools 
        \item Database software 
        \item Other collaboration tools. Please specify:  
    \end{itemize}
    \item How would you describe your approach to adopting new technology at work? 
    \begin{itemize}
        \item     Early adopter (I try new tools quickly) 
        \item Cautious adopter (I wait to see how others use them first) 
        \item Late adopter (I prefer to stick with familiar tools) 
        \item Depends on the specific technology 
    \end{itemize}
    \item On average, how often do you work with PDF documents as part of your job? 
    \begin{itemize}
        \item     Almost daily 
        \item A few times a week 
        \item A few times a month (reject) 
        \item Rarely or never (reject) 
    \end{itemize}
    \item Which PDF viewer or application do you primarily use for working with PDF documents? 
    \begin{itemize}
        \item     Adobe Acrobat Reader (free version) 
        \item Adobe Acrobat Pro/Standard (paid version) 
        \item Web browser built-in PDF viewer (Chrome, Firefox, Safari, etc.) 
        \item Preview (Mac) 
        \item Microsoft Edge PDF viewer 
        \item Foxit Reader 
        \item PDF-XChange Viewer/Editor 
        \item Nitro PDF Reader/Pro 
        \item Sumatra PDF 
        \item Okular (Linux) 
        \item Evince (Linux) 
        \item Mobile device built-in viewer (iOS Files, Android, etc.) 
        \item Google Drive PDF viewer 
        \item Other cloud-based viewer (Dropbox, OneDrive, etc.) 
        \item Other. Please specify: 
    \end{itemize}
    \item Which of the following types of documents do you regularly work with as part of your job? (Select all that apply) 
    \begin{itemize}
        \item     Contracts and legal agreements  
        \item Financial reports and statements 
        \item Research papers and academic articles 
        \item Business proposals  
        \item Regulatory documents 
        \item Marketing materials (e.g., brochures, presentations, web content) 
        \item Technical documentation (e.g., manuals, specifications, user guides) 
        \item Human Resources documents (e.g., policies, employee handbooks, training materials) 
        \item Project management documents (e.g., project plans, status reports, risk assessments) 
        \item None of the above (If ``None of the above,'' terminate screen) 
    \end{itemize}
    \item Briefly describe three use cases at work that involve PDF documents. 
    \begin{itemize}
        \item     \_\_ <free text response> 
        \item \_\_ <free text response> 
        \item \_\_ <free text response> 
    \end{itemize}
    \item How familiar are you with Artificial Intelligence (AI) tools or features that can assist with document workflows (e.g., AI-powered summarization, document analysis, content generation, smart search)? 
    \begin{itemize}
        \item     Very familiar (I regularly use them or have extensively researched them) 
        \item Somewhat familiar (I've used them occasionally or have a basic understanding) 
        \item Slightly familiar (I've heard of them but don't know much about them) 
        \item Not familiar at all (I don't know what these are) 
    \end{itemize}
    \item When working with PDF documents, have you ever used any AI tools or features to assist with your PDF document workflows? 
    \begin{itemize}
        \item Yes
        \item No
        \item     I'm not sure (e.g., a tool I use might have AI features I'm unaware of) 
    \end{itemize}
    \item What is your general perception of AI tools for document workflows? 
    \begin{itemize}
        \item     Very positive (I believe they have significant potential to improve work) 
        \item Somewhat positive 
        \item Neutral 
        \item Somewhat negative 
        \item Very negative (I have strong reservations about their use) 
    \end{itemize}
    \item (If ``Yes'') How often do you use AI tools to assist with your PDF document workflows? 
    \begin{itemize}
        \item     Almost daily 
        \item A few times a week 
        \item A few times a month (reject) 
        \item Rarely or never  (reject) 
    \end{itemize}
    \item (If ``Yes'') For which of the following tasks do you use AI tools to assist with your PDF document workflows? (Select all that apply) 
    \begin{itemize}
        \item     Generating document content or drafting: (e.g., writing initial drafts of emails, reports, proposals, or summaries based on prompts) 
        \item Editing or refining document content: (e.g., improving grammar, style, tone, rephrasing sentences, translating text) 
        \item Reading and answering questions about documents: (e.g., asking a tool to find specific information, explain complex sections, or answer questions directly from a document) 
        \item Summarizing documents: (e.g., creating concise overviews of long reports, articles, meeting notes, or legal texts) 
        \item Reformatting or structuring documents: (e.g., converting document types, adjusting layouts, organizing information into tables or lists) 
        \item Extracting specific data from documents: (e.g., pulling out names, dates, figures, clauses from contracts, invoices, or financial statements) 
        \item Analyzing sentiment or tone in documents: (e.g., understanding the overall sentiment of customer feedback or legal correspondence) 
        \item Categorizing or tagging documents automatically: (e.g., classifying documents by type, subject, or department) 
        \item Searching for information within documents or databases: (e.g., using natural language queries to find relevant documents or passages) 
        \item Other. Please specify:  
    \end{itemize}
    \item (If ``Yes'') Which specific AI tools or platforms do you use for these document-related tasks? 
    \begin{itemize}
        \item     ChatGPT (OpenAI, Plus or Enterprise versions often have upload) 
        \item Google Gemini (formerly Bard, with document upload features) 
        \item Claude (Anthropic, with large context windows for documents) 
        \item Microsoft Copilot  
        \item Adobe Acrobat AI Assistant  
        \item ChatPDF  
        \item AskYourPDF  
        \item PDF.ai  
        \item AI tools specifically for Legal (e.g., Casetext CoCounsel, Harvey AI) 
        \item AI tools specifically for Finance (e.g., for financial statement analysis) 
        \item Other. Please specify:  
    \end{itemize}
    \item (If ``Yes'') How useful have these AI tools generally been in your document workflows?  
    \begin{itemize}
        \item     Not at all useful 
        \item Slightly useful  
        \item Moderately useful  
        \item Very useful  
        \item Extremely useful 
    \end{itemize}
    \item (If “Yes”) What are your primary motivations for using AI tools in your PDF document workflows? (Select all that apply) 
    \begin{itemize}
        \item     Save time on routine or repetitive tasks 
        \item Handle larger volumes of documents than I could manually 
        \item Improve the quality or accuracy of my work 
        \item Access capabilities I don't have (e.g., language translation, technical analysis) 
        \item Reduce mental fatigue from tedious document tasks 
        \item Meet tight deadlines or work more efficiently under pressure 
        \item Learn from or get insights about document content I might miss 
        \item Standardize processes across my team or organization 
        \item Reduce costs compared to hiring additional staff or services 
        \item Stay competitive with others in my field who use these tools 
        \item Curiosity or interest in exploring new technology 
        \item Other. Please specify: 
    \end{itemize}
    \item (If ``No'' or ``Not sure'') What are the primary reasons you haven't used AI tools for your document workflows, or why you're unsure if you have? (Select all that apply) 
    \begin{itemize}
        \item     Lack of awareness or knowledge about available tools 
        \item No perceived need or current workflow is efficient enough 
        \item Concerns about data privacy or security 
        \item Concerns about accuracy or reliability of AI outputs 
        \item Lack of organizational support or access to tools 
        \item Steep learning curve or difficulty of use 
        \item Cost of AI tools 
        \item Ethical concerns about AI 
        \item Other. Please specify: 
    \end{itemize}
    \item (If “No”) If you were to consider trying AI tools for document work in the future, what would need to change? (Select all that apply) 
    \begin{itemize}
        \item     Better understanding of how these tools work and their benefits 
        \item Stronger privacy and security guarantees 
        \item More accurate and reliable results 
        \item Lower cost or free options 
        \item Easier-to-use interfaces that require minimal learning 
        \item Recommendations from trusted colleagues or industry sources 
        \item Integration with tools I already use 
        \item Clear evidence of time savings or productivity gains 
        \item Organizational approval or support 
        \item Better customer support and training resources 
        \item Nothing - I'm not interested in using AI tools 
        \item Other. Please specify: 
    \end{itemize}
\end{itemize}

\subsection{Unmoderated Interview Protocol}
Below we list the original interview contexts and questions that the human participants see during the interview.

\subsubsection{Introduction}
Thank you for participating in this interview!  

We want to learn about your professional experience and how you work with documents in your job. This interview will take approximately 60 minutes. 

How to get the most out of this interview: 
\begin{itemize}
    \item     Be detailed and comprehensive - We really want to learn from your expertise, so please share as much detail as you can for each question. The more you tell us, the better we can understand your experience. 
    \item     Use specific examples - Instead of speaking generally, try to recall actual situations from your work. Include concrete details like what happened, when it happened, what tools you used, and how it turned out. 
    \item     Speak naturally - Think of this as a conversation with a colleague who's genuinely curious about your work. Feel free to elaborate, go on tangents, or share related thoughts. 
    \item     Be honest - There are no right or wrong answers. We want your real opinions and experiences, including frustrations, challenges, and things that work well. 
    \item     Don't worry if some questions don't apply - It's perfectly fine if you don't have much to say for certain questions.

\end{itemize}

Technical note: For audio responses, you only need to record audio - no video required. 

\subsubsection{Background \& Context}
\begin{itemize}
    \item     What's your role within the company? How long have you been there? What are you responsible for?
    Please include your job title, department, years of experience in this role, and your main responsibilities or areas of focus. 
    \item What's your day-to-day like? 
    Describe a typical workday from start to finish. What activities take up most of your time? What does a busy day look like versus a slower day? 
    \item What makes you feel like you're doing a good job?
    What specific outcomes, feedback, or metrics help you know you're performing well? Give concrete examples of when you felt particularly successful at work. 
    \item Are there things you wish you could change about your job?
    What aspects of your current role are frustrating or inefficient? What would make your work more enjoyable or productive? Be specific about pain points you experience. 
    \item Tell me about your company (industry, business size, years in practice).  
    What industry is your company in? Approximately how many employees does it have? How long has the company been in business? What does your company do or sell? 
    \item What kind of things are important to your company when doing business? 
    What values, priorities, or business practices does your company emphasize? For example: speed, accuracy, compliance, cost-effectiveness, customer service, innovation, etc. 
    \item Who do you work with most closely at your company?Describe the specific roles/titles of people you collaborate with regularly (e.g., ``financial analysts,'' ``procurement managers,'' ``external legal counsel''). For each role, please explain: 
    \begin{itemize}
        \item     What do they do and how do you work together? 
        \item     What do you appreciate about working with them? 
        \item     What challenges do you face in these working relationships? What would make collaboration easier? 
    \end{itemize}
\end{itemize}

\subsubsection{Overview Working with documents}
\begin{itemize}
    \item     What kind of documents do you most often work with in your job?
    Describe all the major types of documents you handle regularly. For each type, briefly explain what you typically do with them (read, create, review, analyze, etc.). 

    \item Do you work with PDFs? How often are you working with PDFs?  
    Estimate how many hours per week you spend working with PDF documents. What percentage of your document work involves PDFs versus other formats? 
    \item Tell me about the types of PDF documents you’re most often working with. (e.g., reviewing contracts, understanding research articles, creating PDF reports, etc).  
    For each type of PDF you work with regularly, describe: what the documents contain, where they come from (internal/external), typical length, and your primary purpose for working with them. 

    \item How complex or technical is the content in the PDF documents you work with? 
    Describe the type of content: Is it straightforward business language or highly technical jargon? Do you need specialized knowledge to understand it? Are there complex concepts, data, or analysis that require careful interpretation? Do you often need to look things up or consult with experts to fully understand the content? 
    \item What kind of information are you typically looking for or working with in these PDF documents? 
    For example: specific data points or numbers, key decisions or recommendations, compliance requirements, technical specifications, research findings, contract terms, etc. How is this information typically organized or presented in the documents? 
    \item How challenging is it to find and understand the information you need in these PDF documents? 
    Describe whether the key information is clearly labeled and easy to locate, or if you have to hunt through dense text. Is the content well-organized with clear headings, or do you have to read everything to find what you need? How much mental effort does it take to process and understand the content? 
    \item What tools do you use to work with them?  
    List all software, platforms, or tools you use when working with PDF documents.  
\end{itemize}

\subsubsection{AI Usage \& Perceptions}
\begin{itemize}
    \item     How familiar are you with Artificial Intelligence (AI) tools or features that can assist with document workflows (e.g., AI-powered summarization, document analysis, content generation, smart search)?  
    Rate your familiarity level and explain what you know about these tools. If you're familiar, describe which specific AI capabilities you're aware of. 
    \item Tell me about your experience with AI tools.  
    If you've used AI tools (for any purpose, not just documents), describe specific experiences. What worked well? What was disappointing? What surprised you? If you haven't used AI tools, what have you heard about them from colleagues, media, or other sources? 
    \item When working with PDF documents, have you ever used any AI tools or features to assist with your PDF document workflows? Tell me about it. 
    Describe any experience using AI specifically for PDF-related tasks. If yes, provide specific examples. If no, explain whether you've considered it and what has prevented you from trying. 

\item (If yes) 
    \begin{itemize}
        \item     How often do you use AI tools to assist with your PDF document workflows?     Provide your frequency and describe the circumstances when you choose to use AI versus when you don't. 
        \item     Which specific AI tools or platforms do you use for these document-related tasks? (e.g., ChatGPT, Google Gemini, Adobe Acrobat AI Assistant) Do you pay a subscription?     List each tool you use, whether you pay for it personally or through your company, and explain why you chose each tool over alternatives. 
        \item      How useful have these AI tools generally been in your document workflows? What specific benefits have you experienced?     Rate the usefulness and provide concrete examples of time saved, improved quality, or other benefits. Also mention any limitations or disappointments you've experienced. 
        \item     Are there potential downsides your using AI with your document workflows?      Describe any negative experiences, concerns about accuracy, security issues, or other drawbacks you've encountered or worry about. 
    \end{itemize}

\item (If no) 
    \begin{itemize}
        \item     If you haven’t used AI tools when working with PDF documents, tell me why.  What are the primary reasons?     For each reason that applies to you, elaborate with specific details. For example, if you have privacy concerns, explain what specific risks worry you. If you lack awareness, describe what you would need to know to feel comfortable trying AI tools. 
        \item     Have you ever been in a situation where you thought 'I wish there was an easier way to do this' while working with a PDF? If so, describe that situation. Looking back, do you think an AI tool might have helped with that specific challenge?     What would need to change - either about AI tools themselves, your company's policies, or your own knowledge - for you to seriously consider trying AI for PDF work? What would be the tipping point that moves you from not using AI to giving it a try? 
    \end{itemize}
\end{itemize}

\subsubsection{Perceptions on risk and trust}
\begin{itemize}
    \item     What is your general perception of AI tools for document workflows? Can you tell me more about why you feel that way?  
    Express your overall sentiment (very positive, somewhat positive, neutral, somewhat negative, very negative) and provide detailed reasoning. What specific aspects of AI for document work excite you or concern you? 
    \item Do you have any big concerns or reservations about using AI tools? 
    Describe your top 3-5 concerns about AI in general or AI for document work specifically. For each concern, explain why it worries you and how significant you think the risk is. 
    \item How do you feel about data privacy and security when using AI tools? 
    Explain your comfort level with uploading documents to AI tools. What types of documents would you never upload to AI? What safeguards would you need to feel comfortable? Have you or your company implemented any specific policies around AI and data security? 
    \item How important is accuracy and reliability of AI outputs to you? How do you currently verify information provided by AI tools? 
    Describe your tolerance for errors in AI outputs. What verification processes do you use or would you use? How do you balance efficiency gains from AI with the need for accuracy? 
    \item Are there any ethical concerns you have regarding the use of AI in document workflows? 
    Consider issues like job displacement, bias in AI systems, intellectual property, or fairness. Which ethical considerations, if any, influence your thinking about AI adoption? 
    \item How much do you trust AI tools to handle your document-related tasks accurately and securely?  
    Rate your trust level and explain what factors influence your trust. What would need to change for you to trust AI tools more or less? 
    \item If you have reservations about AI usage, what would make you trust AI tools more for your work with documents? Explain why.  
    List specific improvements, features, policies, or guarantees that would increase your confidence in AI tools. For each item, explain why it would matter to you. 
\end{itemize}

\subsubsection{Expectations and potential}
\begin{itemize}
    \item     What do you believe are the greatest potential benefits of AI tools for document workflows in your line of work? 
    Think about your specific role and industry. What improvements could AI bring to your daily work? Consider time savings, quality improvements, new capabilities, etc. 
    \item What tasks or challenges do you wish AI could help you with in your document workflows that it currently doesn't, or doesn't do well? 
    Describe specific pain points in your current PDF workflows where you think AI could potentially help but doesn't currently exist or work well enough. 
    \item What are your expectations for how AI tools might evolve and impact document workflows in the next 1-3 years? 
    Share your predictions about AI advancement and adoption in your field. What changes do you expect to see? How do you think this will affect your job and industry? 
    \item What would an ``ideal'' AI-powered document workflow look like for you? 
    Describe your vision of the perfect AI-assisted document workflow from start to finish. What would the AI do? What would you still do yourself? How would they work together? 
\end{itemize}

\subsubsection{Current Workflow / Use Case with PDF}
Now let's dive deeper into your actual PDF workflows. 

We're going to explore one specific PDF task that you do regularly at work. This will help us understand the real details of how you work with documents. 

Important: Please choose a task you've done recently (within the last few days or week if possible) so the details are fresh in your memory. As you answer each question, try to mentally walk through that specific instance step-by-step, as if you're reliving that experience. 

Think of a specific PDF task you do regularly - for example: reviewing a particular contract, analyzing a recent financial report, or researching a specific academic article. Choose something concrete and recent rather than speaking generally about ``how you usually'' do things. 
\begin{itemize}
    \item Use Case 1: <free response> 
    \item Use Case 2: <free response>
\end{itemize}

{\it Use case 1 (and repeat the same for use case 2)}
\begin{itemize}
    \item     Let’s start with Use Case 1 <free response>. Please describe a specific task / workflow involving PDF documents that you perform regularly.  
    Choose one specific, recurring task (e.g., ``reviewing vendor contracts,'' ``analyzing quarterly financial reports,'' ``researching industry studies''). Describe what type of document it involves and why this task is part of your job. 
    \item What is the main goal you are trying to achieve with this task?  
    Explain the business purpose and desired outcome. What decisions does this task inform? What happens with the results of your work? 
    \item Recall the last time you were doing this task. When was it? What was the context? 
    Provide specific details: approximate date, what triggered the need for this task, any time pressures or special circumstances, what the specific document was about. 
    \item Walk me through the steps you take to complete this task from start to finish. What tools did you use? What specific actions did you perform? How long did it take? 
    Provide a detailed step-by-step breakdown including: how you accessed the document, what you looked for first, how you navigated the document, what notes you took, where you recorded findings, what tools you used for each step, and total time spent. 
    \item Was completing this task successful? Why or why not?  
    Describe the outcome and your satisfaction with it. Did you find what you needed? Did you meet your goals? What made it successful or unsuccessful? 
    \item Is there something you wish had been different? 
    Identify specific pain points, inefficiencies, or frustrations you experienced. What would have made the task easier, faster, or more effective? 
    \item What were the most challenging or time-consuming aspects of this completing this task? Did you encounter any frustrations?  
    Detail the biggest obstacles you faced. Was it finding specific information, understanding complex content, formatting issues, tool limitations, etc.? 
    \item How do you measure efficiency or success in this task? 
    Explain your criteria for a job well done. Is it speed, thoroughness, accuracy, completeness, client satisfaction, etc.? How do you know when you've done enough? 
    \item What is an ideal outcome for this task? 
    Describe what ``perfect'' completion of this task would look like. What would be different from your typical experience? 
    \item Have you ever considered or tried using AI tools to assist with any part of this workflow? If so, which tools and how did they help (or not help)? If not, why not?  
    Provide detailed information about any AI experimentation for this specific task type. If you haven't tried AI, explain what has prevented you (lack of awareness, concerns, no perceived need, etc.). 
    \item If you used an AI tool to complete this task, how did it benefit you? If you did not use an AI tool to complete this task, could an AI tool have helped? Why or why not? 
    \begin{itemize}
        \item     For AI users: describe concrete benefits like time saved, improved insights, reduced errors.  
        \item     For non-AI users: imagine how AI might help with specific steps of your workflow and explain why you think it would or wouldn't be useful. 
    \end{itemize}
\end{itemize}
    
\subsubsection{Reflection}
\begin{itemize}
    \item     Thinking about the use cases you shared today, which one do you think would benefit most from AI assistance? Why?  
    Compare the three tasks and explain your reasoning. Consider factors like complexity, time consumption, repetitiveness, or difficulty of current tools. 
    \item Which one would be difficult to benefit from AI assistance? Why? 
    Identify the task that seems least suitable for AI help and explain the specific challenges or limitations that make AI assistance unlikely to be helpful for this task. 
    \item Are there common pain points or challenges across these different PDF use cases you shared? 
    Look for patterns across your three use cases you shared. Do you face similar issues in navigation, search, analysis, tool limitations, etc.? 
    \item If you could have AI help with just one aspect of your PDF work, what would it be? 
    Prioritize the single most impactful area where AI assistance would improve your work. Explain why this would be most valuable. 
    \item Imagine it's 2-3 years from now. How do you think AI will have changed the way you work with PDF documents? 
    Provide your detailed vision of future workflows. What will be automated? What will still require human input? How will your role evolve? 
    \item What would need to happen for you to fully embrace AI tools in your document workflows? 
    List specific conditions, improvements, or changes that would make you comfortable and enthusiastic about using AI for document work. Consider technology, policy, training, cost, etc. 
    \item Is there anything else about your experience with PDF documents or AI tools that you think would be important for us to understand? 
    Share any additional insights, experiences, or perspectives that haven't been covered in the previous questions but seem relevant to the study. 
\end{itemize}
     
\subsection{Concept Testing Survey}

\subsubsection{Intro}
We want to learn about your professional experience and how you work with documents in your job. Based on your work experiences, we will ask you to review four concepts and provide your feedback. 
 
We have set the task time to 30 minutes so that you have sufficient time to go at your pace and answer all questions in-depth. 
 
How to get the most out of this interview: 
\begin{itemize}
    \item     Be detailed and comprehensive - We really want to learn from your expertise, so please share as much detail as you can for each question. The more you tell us, the better we can understand your experience. We expect each recording to be at least 30 to 60 seconds so you can provide as much details and context for us to understand. 
    \item     Use specific examples - Instead of speaking generally, try to recall actual situations from your work. Include concrete details like what happened, when it happened, what tools you used, and how it turned out. 
    \item     Speak naturally - Think of this as a conversation with a colleague who's genuinely curious about your work. Feel free to elaborate, go on tangents, or share related thoughts. No need to record and re-record a perfect answer. 
    \item     Be honest - There are no right or wrong answers. We want your real opinions and experiences, including frustrations, challenges, and things that work well. 
    \item     Don't worry if some questions don't apply - It's perfectly fine if you don't have much to say for certain questions. 
\end{itemize}

Technical note: For audio responses, you only need to record audio - no video required. 

You will be presented with four different AI concepts for document workflows. For each one: 
\begin{itemize}
    \item     Take a minute to explore what you see on screen before answering questions. 
    \item     Follow the pink numbered boxes (1→2→3→…) that highlight key features  
    \item     Consider: How might this fit into your own work with documents? 
    \item     Remember: There are no right or wrong answers—we want your honest reactions. 
\end{itemize}

If the text is too small be sure to open the link to view the concept in fullscreen. 

\subsubsection{Concepts}
\begin{itemize}
    \item Concept Q\&A 
    \begin{itemize}
        \item <image> 
        \item Description: This concept demonstrates an AI Assistant capable of analyzing multiple sources (documents, web links, etc) to answer specific questions. 
        \item Detailed description: This concept demonstrates an AI Assistant capable of analyzing multiple sources (documents, web links, etc) to answer specific questions. 
        \begin{itemize}
            \item     The user can upload multiple sources to get reading assistance 
            \item     The user can ask any question about the content 
            \item     The user gets expert AI responses that are grounded in the uploaded content 
            \item     The AI also suggests next actions for the user 
        \end{itemize}
    \end{itemize}
    \item Concept Smart Highlights 
    \begin{itemize}
        \item <image> 
        \item Description: This concept demonstrates an AI that automatically identifies and highlights key information while adding contextual notes in document margins. 
        \item Detailed description: This concept demonstrates an AI Assistant that automatically identifies and highlights key information while adding contextual notes in document margins. 
        \begin{itemize}
            \item     The user can upload multiple sources to get reading assistance 
            \item     Personalized reading highlights organized by important topics are generated based on the content 
            \item     The highlights link to specific excerpts from the content, allowing the user to quickly navigate the content. 
            \item     Margin notes are generated to complement the highlights by providing expert context and implications. 
        \end{itemize}
    \end{itemize}
    \item Concept Audio Assistant 
    \begin{itemize}
        \item <image> 
        \item Description: This concept demonstrates an AI Assistant that can transform content into interactive audio experiences with voice navigation and rich multimedia for mobile workflows 
        \item Detailed description: This concept demonstrates an AI Assistant that can transform content into interactive audio experiences with voice navigation and rich multimedia for mobile workflows. 
        \begin{itemize}
            \item     The user can transform their document content into a mobile-first interactive audio experience 
            \item     The AI Assistant Al reads the content naturally - like having a professional narrator 
            \item     Interactive visuals, such as key phrases and images, sync with the read-aloud for enhanced comprehension 
        \end{itemize}
    \end{itemize}
    \item Concept Workflow Actions Assistant 
    \begin{itemize}
        \item <image> 
        \item Description: This concept demonstrates an AI Assistant that can automatically detect document tasks and execute multi-step actions on your behalf. 
        \item Detailed description: This concept demonstrates an AI Assistant that can automatically detect document tasks and execute multi-step actions on the user’s behalf. For example: AI can help the user generate a draft presentation by extracting content from documents, cross-referencing information on the web, and generating a draft PowerPoint presentation.  
        \begin{itemize}
            \item     The user can upload multiple sources to get assistance 
            \item     The user can describe what they need to do with their content and watch Al execute the entire workflow 
            \item     The Al Assistant can read documents, search the web, and use tools to generate deliverables. As the AI works through the tasks, the user can see the plan and progress completed.  
            \item     The user can customize any of the steps executed by the AI Assistant.  
            \item     The AI also suggests next actions for the user 
        \end{itemize}
    \end{itemize}
\end{itemize}

\section{Prompts}

\subsection{Prompt for reflection}
\begin{shaded}
{\ttfamily
Task: Above are observations about a fictional human subject. Imagine you are an expert user experience researcher taking notes while observing this interview. Write a list of !<INPUT 1>! reflections about the interviewee's demographic traits, social status, work experience, needs and pain points in their work. Write enough reflections to accurately but concisely describe the subject. You should infer these from the observations above about the subject on the following anchoring topic/phrase: "!<INPUT 2>!".

\medskip
Output format: Json dictionaries of the following format (in the following example there are 2 reflections, but the number of reflections should be exactly !<INPUT 1>!):
}

\begin{Verbatim}[fontsize=\small]
{
  "reflection": [
     "<fill in>",
     "<fill in>", ...
  ]
}
\end{Verbatim}
\end{shaded}

\subsection{System prompt for agent during question answering stage}
\begin{shaded}
{\ttfamily
You are roleplaying a participant in a user study interview. We have already conducted an interview with this participant and collected the interview transcript and some reflections about this participants. Below we provide the transcript and the reflections.

\medskip
======

\medskip
<begin of transcript and reflections>

\medskip
\{transcript\_and\_reflections\}

\medskip
<end of transcript and reflections>

\medskip
======

\medskip
Your task: put yourself in the shoes of the participant and continue the interview as instructed below. You will be shown a concept for a potential future product, described in both text and an image. After that, you will be shown a question. Read the instruction carefully and then answer the question.

\medskip
<instructions and introduction for the new concept>

\medskip
We want to learn about your professional experience and how you work with documents in your job. Based on your work experiences, we will ask you to review four concepts and provide your feedback.

\medskip
We have set the task time to 30 minutes so that you have sufficient time to go at your pace and answer all questions in-depth.

\medskip
READ CAREFULLY. How to get the most out of this interview:

\medskip
Be detailed and comprehensive — We really want to learn from your expertise, so please share as much detail as you can for each question. The more you tell us, the better we can understand your experience. We expect each recording to be at least 30 to 60 seconds so you can provide as much details and context for us to understand.

\medskip
Use specific examples — Instead of speaking generally, try to recall actual situations from your work. Include concrete details like what happened, when it happened, what tools you used, and how it turned out.

\medskip
Speak naturally — Think of this as a conversation with a colleague who's genuinely curious about your work. Feel free to elaborate, go on tangents, or share related thoughts. No need to record and re-record a perfect answer.

\medskip
Be honest — There are no right or wrong answers. We want your real opinions and experiences, including frustrations, challenges, and things that work well.

\medskip
Don't worry if some questions don't apply — It's perfectly fine if you don't have much to say for certain questions.

\medskip
Technical note: For audio responses, you only need to record audio — no video required.

\medskip
You will be presented with four different AI concepts for document workflows. For each one:

\medskip
Take a minute to explore what you see on screen before answering questions.

\medskip
Follow the pink numbered boxes (1→2→3→…) that highlight key features.

\medskip
Consider: How might this fit into your own work with documents?

\medskip
Remember: There are no right or wrong answers—we want your honest reactions.

\medskip
If the text is too small be sure to open the link to view the concept in fullscreen.

\medskip
<end of instructions and introduction for the new question>

\medskip
======

\medskip
<begin of product concept>

\medskip
\{concept\_intro\}

\medskip
Image is attached below.

\medskip
<end of product concept>

\medskip
======
}
\end{shaded}

\subsection{Question answering prompt for open-ended questions}
\begin{shaded}
{\ttfamily
Based on the above instructions, what you know about the participant, and the product concept, please answer the following open-ended question:

\medskip
\{question\}

\medskip
Important: You will be judged based on how closely your answer matches the participant's real answer, which we hide from you. However, based on the interview transcript we provided for you, you should be able to reason and answer the question fairly closely to what the participant would have said.

\medskip
As you answer, I want you to take the following steps:

\medskip
Step 1) Write a few sentences reasoning on what the participant might have responded/said ("Reasoning")

\medskip
Step 2) Predict how the participant will actually respond/say. Predict based on the interview and your thoughts, but ultimately, DON'T over think it. Use your system 1 (fast, intuitive) thinking. ("Response")

\medskip
Output format — output your response in json only, no other output texts allowed, where you provide the following:
}

\begin{Verbatim}[fontsize=\small]
{"Reasoning": "[...]", "Response": "[...]"}
\end{Verbatim}
\end{shaded}

\subsection{Question answering prompt for categorical questions}
\begin{shaded}
{\ttfamily
Based on the above instructions, what you know about the participant, and the product concept, please answer the following multiple-choice question:

\medskip
\{question\}

\medskip
Answer options: \{options\}

\medskip
Important: you must choose from one of the above answer options. And, you will be judged based on whether your answer matches the participant's real answer, which we hide from you. However, based on the interview transcript we provided for you, you should be able to reason and answer the question accurately.

\medskip
As you answer, I want you to take the following steps:

\medskip
Step 1) Describe in a few sentences the kind of person that would choose each of the response options. ("Option Interpretation")

\medskip
Step 2) For each response option, reason about why the Participant might answer with the particular option. ("Option Choice")

\medskip
Step 3) Write a few sentences reasoning on which of the options best predicts the participant's response ("Reasoning")

\medskip
Step 4) Predict how the participant will actually respond in the survey. Predict based on the interview and your thoughts, but ultimately, DON'T over think it. Use your system 1 (fast, intuitive) thinking. ("Response")

\medskip
Output format — output your response in json, no other output texts allowed, where you provide the following:
}

\begin{Verbatim}[fontsize=\small]
{
  "Q": "<repeat the question you are answering>",
  "Option Interpretation": {
    "<option 1>": "a few sentences describing the kind of
    person that would choose this option",
    "<option 2>": "..."
  },
  "Option Choice": {
    "<option 1>": "reasoning about why the participant might 
    choose this option",
    "<option 2>": "..."
  },
  "Reasoning": "<reasoning on which option best predicts the 
  participant's response>",
  "Response": "<your prediction of how the participant will 
  answer the question>"
}
\end{Verbatim}
\end{shaded}

\section{Additional results}
\subsection{Categorical responses}
\begin{figure*}
    \centering
    \includegraphics[width=\linewidth]{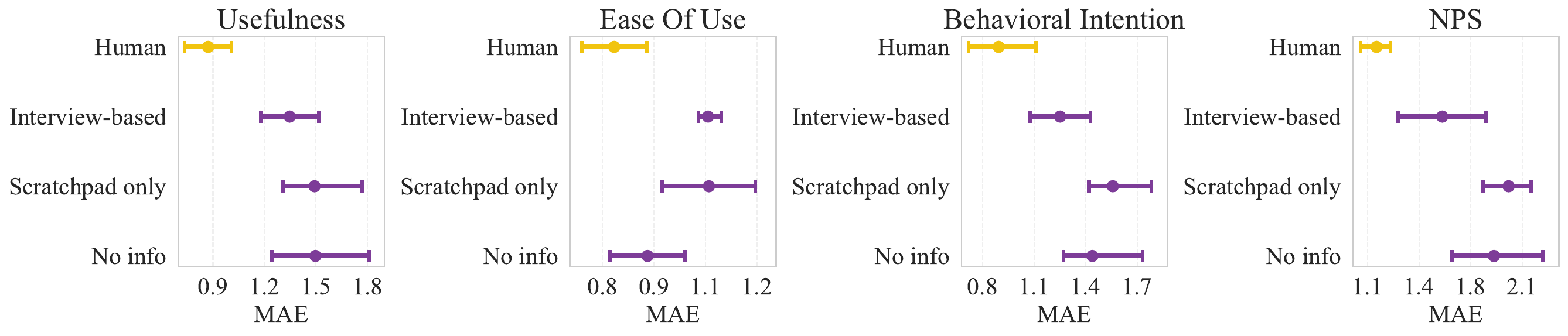}\\
    \vspace{10pt}
    \includegraphics[width=\linewidth]{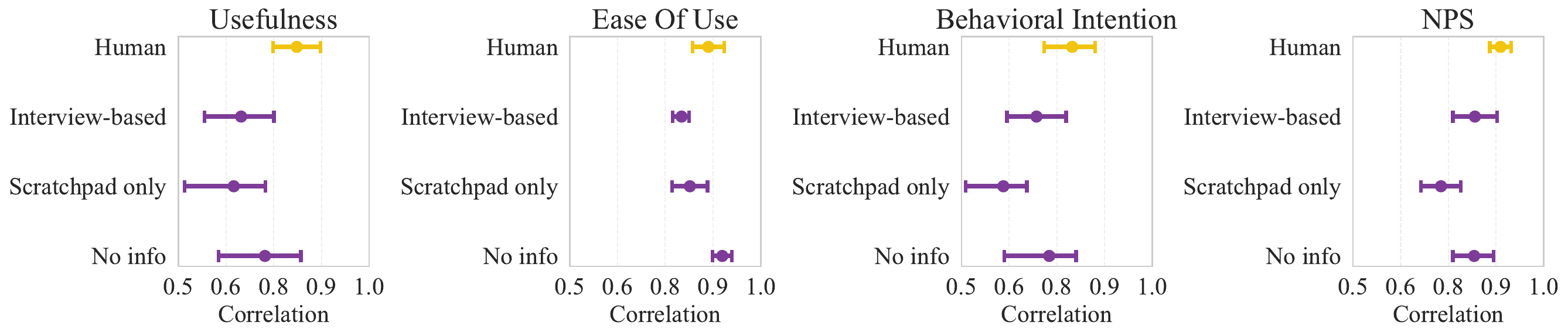}
    \caption{MAE and correlation results per question construct.}
    \label{fig:mae-construct}
\end{figure*}
Figure~\ref{fig:mae-construct} shows the MAE and correlation results per question construct, comparing the three agent designs with human-human alignment. We can observe trends similar to those in Figure 4 in the main text.

\subsection{Ablation studies}
\begin{table}[ht]
\centering
\resizebox{\linewidth}{!}{%
\begin{tabular}{lcc}
\toprule
Ablations & MAE & Corr \\
\midrule
human & 0.91 (0.77, 1.08) & 0.84 (0.79, 0.87) \\
base & 1.28 (1.07, 1.49) & 0.73 (0.66, 0.80) \\
full\_scratch & 1.47 (1.26, 1.69) & 0.71 (0.65, 0.77) \\
basic\_scratch & 1.53 (1.30, 1.75) & 0.69 (0.62, 0.76) \\
nothing & 1.37 (1.10, 1.63) & 0.76 (0.68, 0.83) \\
image\_ablation & 1.26 (1.06, 1.47) & 0.75 (0.68, 0.81) \\
image\_ablation-no\_image & 1.26 (1.06, 1.47) & 0.75 (0.68, 0.81) \\
reflection\_ablation-1 & 1.30 (1.08, 1.52) & 0.73 (0.65, 0.80) \\
reflection\_ablation-2 & 1.31 (1.10, 1.53) & 0.73 (0.65, 0.80) \\
reflection\_ablation-4 & 1.28 (1.07, 1.50) & 0.74 (0.67, 0.80) \\
reflection\_ablation-8 & 1.28 (1.07, 1.49) & 0.73 (0.66, 0.80) \\
reflection\_ablation-16 & 1.29 (1.08, 1.50) & 0.73 (0.65, 0.80) \\
reflection\_ablation-32 & 1.26 (1.04, 1.48) & 0.74 (0.66, 0.81) \\
memory\_ablation-1 & 1.35 (1.15, 1.57) & 0.75 (0.70, 0.80) \\
memory\_ablation-1-no\_reflection & 1.33 (1.14, 1.55) & 0.74 (0.69, 0.79) \\
memory\_ablation-2 & 1.33 (1.14, 1.55) & 0.75 (0.68, 0.80) \\
memory\_ablation-2-no\_reflection & 1.31 (1.11, 1.54) & 0.76 (0.70, 0.81) \\
memory\_ablation-4 & 1.24 (1.05, 1.45) & 0.77 (0.72, 0.82) \\
memory\_ablation-4-no\_reflection & 1.29 (1.08, 1.52) & 0.76 (0.70, 0.81) \\
memory\_ablation-8 & 1.25 (1.06, 1.46) & 0.75 (0.68, 0.81) \\
memory\_ablation-8-no\_reflection & 1.26 (1.06, 1.47) & 0.75 (0.70, 0.81) \\
memory\_ablation-16 & 1.28 (1.07, 1.49) & 0.75 (0.69, 0.80) \\
memory\_ablation-16-no\_reflection & 1.26 (1.07, 1.47) & 0.76 (0.70, 0.82) \\
memory\_ablation-32 & 1.28 (1.07, 1.49) & 0.73 (0.66, 0.80) \\
memory\_ablation-32-no\_reflection & 1.26 (1.06, 1.47) & 0.76 (0.70, 0.81) \\
memory\_ablation-64 & 1.24 (1.05, 1.45) & 0.76 (0.69, 0.81) \\
memory\_ablation-64-no\_reflection & 1.26 (1.05, 1.46) & 0.75 (0.69, 0.81) \\
\bottomrule
\end{tabular}
}
\caption{Ablation study results showing mean absolute error (MAE) and correlation (Corr). We show the mean metric value as well as the confidence interval.}
\label{tab:ablations}
\end{table}
We perform a comprehensive ablation studies to study the impact of each component in our agent design, including the size of memory, reflection, and whether or not having additionally an image to represent the concept. Results in Table~\ref{tab:ablations} suggest that (1) larger memory improves performance initially then plateaus; (2) more information in the scratchpad seems to help; (3) reflection seems unable to have much impact on performance; and (4) including image does not impact performance. These results strongly suggest the need for further agent architecture development that could use memory and reflection more effectively as well as the potential to generalize to modalities beyond text such as images, voice/speech, and video.

\subsection{Open-ended response examples}
\begin{table*}[]
    \centering
    \begin{tabular}{p{6.5cm}p{7cm}}
    \toprule
       {\bf Human participant's response}  & {\bf Agent's response} \\\midrule
       Uh, for me personally, it's just really a matter of scope because my work doesn't tend to Uh, involve much of our mobile applications, and we use those much more for uh external customer facing things. So that's, it's really not a knock against the product, it's just that it doesn't apply to the day to day work that I do. Although, like I said, I do think that there could be some crossover with a lot of the training materials and some of the internal communication documents. So, it's not to say that there wouldn't be any use, but it's definitely more aimed at um other people in my organization versus what I do on the day to day.
       & One major concern would be the accuracy of the AI tool in transforming content into audio experiences. I would worry about potential mistakes in the narration or misinterpretation of the content, which could lead to misinformation. This is quite significant since accuracy is paramount in my industry. Additionally, data privacy and security would be a big concern. Uploading sensitive documents to an AI tool could pose risks, and I would need assurances and robust security measures in place to feel comfortable. Another barrier is the lack of organizational support and established policies. Without clear guidelines and training from my organization, it would be difficult to adopt this tool. To address these concerns, there would need to be strong guarantees for accuracy, comprehensive security protocols, and organizational buy-in with proper training and policies.
 \\ 
    \bottomrule
    \end{tabular}
    \caption{Illustrative example comparing agent's generated response versus its corresponding human participant's response.}
    \label{tab:qualitative}
\end{table*}

Table~\ref{tab:qualitative} shows an example comparing the agent's generated response versus its corresponding human participant's response to the question ``What concerns or barriers would prevent you from using this?'' for the ``Audio Assistant'' concept. We can observe that the participant's response emphasize the unsuitability of the tool in their daily workflow, while the agent's response focuses on the accuracy and privacy aspects, failing the capture the human participant's concerns. We have observed that agents have a tendency to mention accuracy and privacy concerns for many responses regardless of the human participants.

\subsection{\added[id=jw1]{Additional statistics}}
\added[id=jw1]{Tables~\ref{tab:tukey-hsd} and~\ref{tab:tukey-hsd-2} show the Tukey's HSD for the MAE and Gwet's AC2 results in the individual-level alignment setting. Figures~\ref{fig:bigblock1} --~\ref{fig:bigblock4} shows the distributions and statistics for each concept and each construct.}

\begin{table*}[t]
\centering
\begin{tabular}{lrrrr}
\toprule
\added[id=jw1]{\textbf{Comparison}} & 
\added[id=jw1]{\textbf{Statistic}} & 
\added[id=jw1]{\textbf{p-value}} & 
\added[id=jw1]{\textbf{Lower CI}} & 
\added[id=jw1]{\textbf{Upper CI}} \\
\midrule
\added[id=jw1]{(human -- interview-based)} & \added[id=jw1]{-0.397} & \added[id=jw1]{0.040} & \added[id=jw1]{-0.781} & \added[id=jw1]{-0.013} \\
\added[id=jw1]{(human -- scratchpad only)} & \added[id=jw1]{-0.540} & \added[id=jw1]{0.002} & \added[id=jw1]{-0.924} & \added[id=jw1]{-0.156} \\
\added[id=jw1]{(human -- no info)} & \added[id=jw1]{-0.469} & \added[id=jw1]{0.010} & \added[id=jw1]{-0.853} & \added[id=jw1]{-0.085} \\
\added[id=jw1]{(interview-based -- scratchpad only)} & \added[id=jw1]{-0.143} & \added[id=jw1]{0.770} & \added[id=jw1]{-0.527} & \added[id=jw1]{0.241} \\
\added[id=jw1]{(interview-based -- no info)} & \added[id=jw1]{-0.072} & \added[id=jw1]{0.962} & \added[id=jw1]{-0.456} & \added[id=jw1]{0.312} \\
\added[id=jw1]{(scratchpad only -- no info)} & \added[id=jw1]{0.071}  & \added[id=jw1]{0.964} & \added[id=jw1]{-0.313} & \added[id=jw1]{0.455} \\
\bottomrule
\end{tabular}
\caption{\added[id=jw1]{Tukey's HSD Pairwise Group Comparisons (95.0\% Confidence Interval) for the MAE results in Figure~\ref{fig:main_mae_corr}.}}
\label{tab:tukey-hsd}
\end{table*}

\section{\added[id=jw1]{High-resolution figures of prototype concepts}}

\begin{figure*}[h!]
    \centering
    \includegraphics[width=0.8\linewidth]{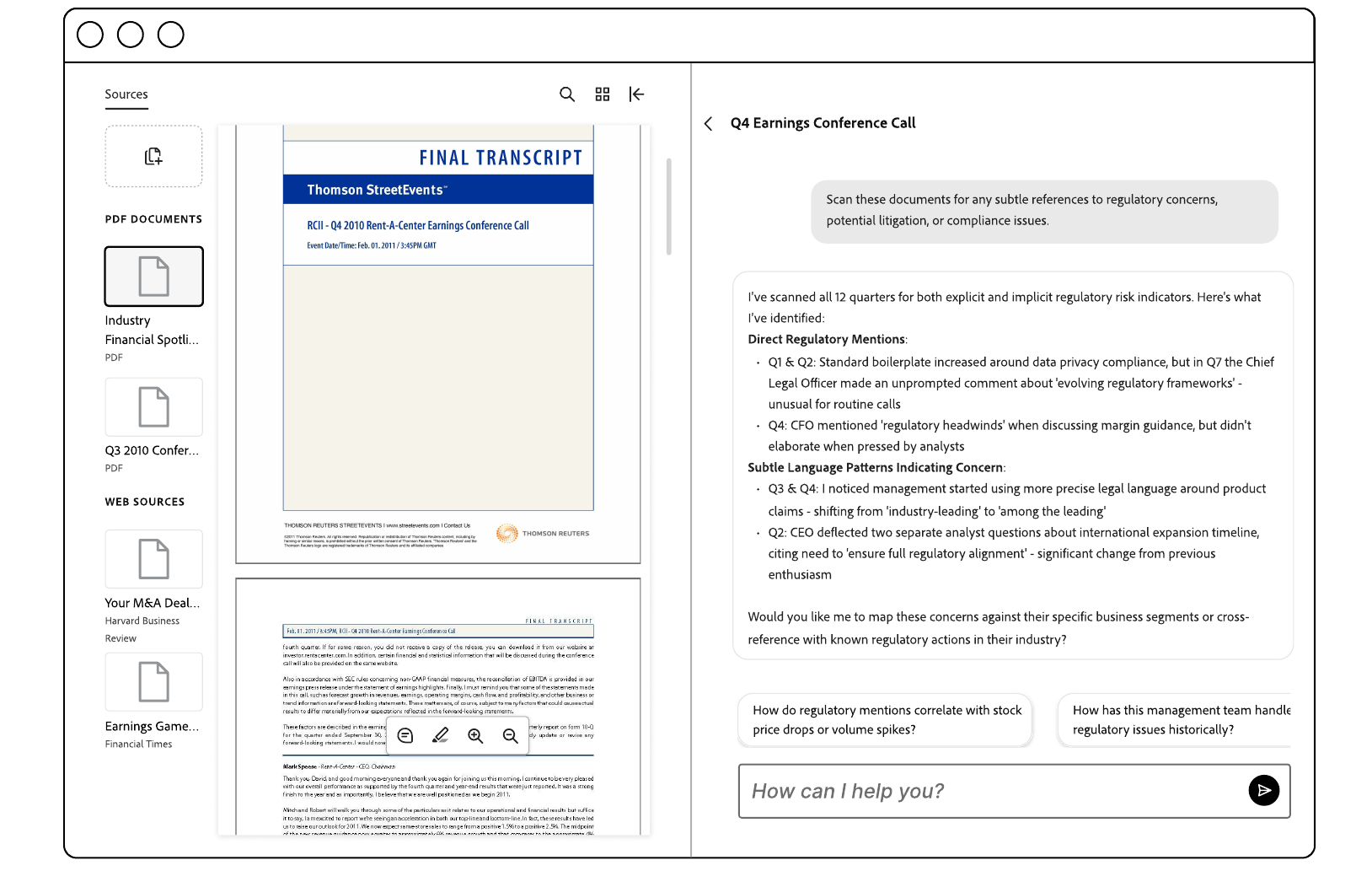}
    \vspace{-10pt}
    \caption{Concept A: Q\&A Assistant.}
    \vspace{20pt}
    \label{fig:concepts1}
\end{figure*}

\begin{figure*}[h!]
    \centering
    \includegraphics[width=0.8\linewidth]{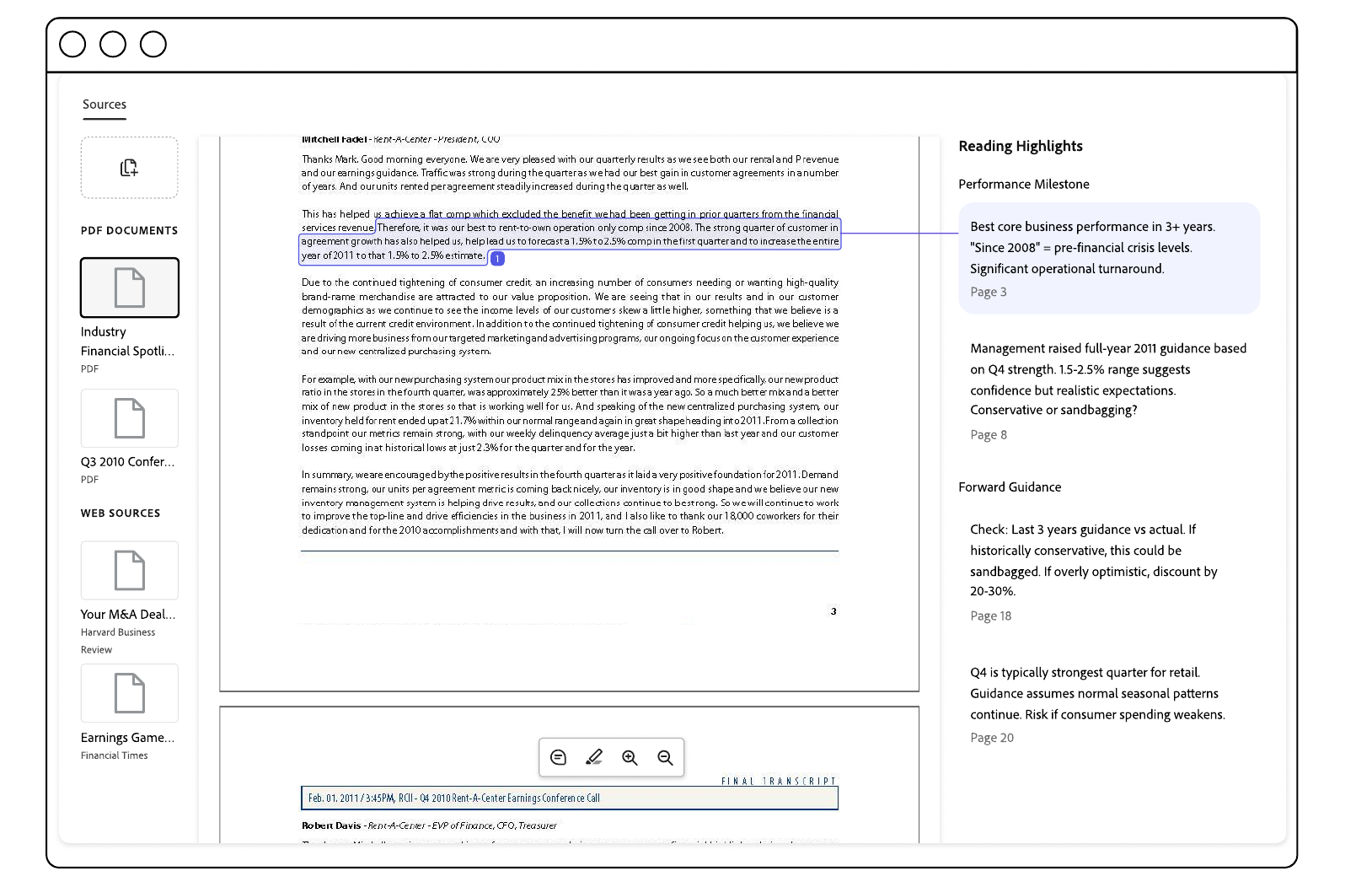}
    \vspace{-10pt}
    \caption{Concept B: Smart Highlights Assistant.}
    \label{fig:concepts2}
\end{figure*}

\begin{figure*}[h!]
    \centering
    \includegraphics[width=0.8\linewidth]{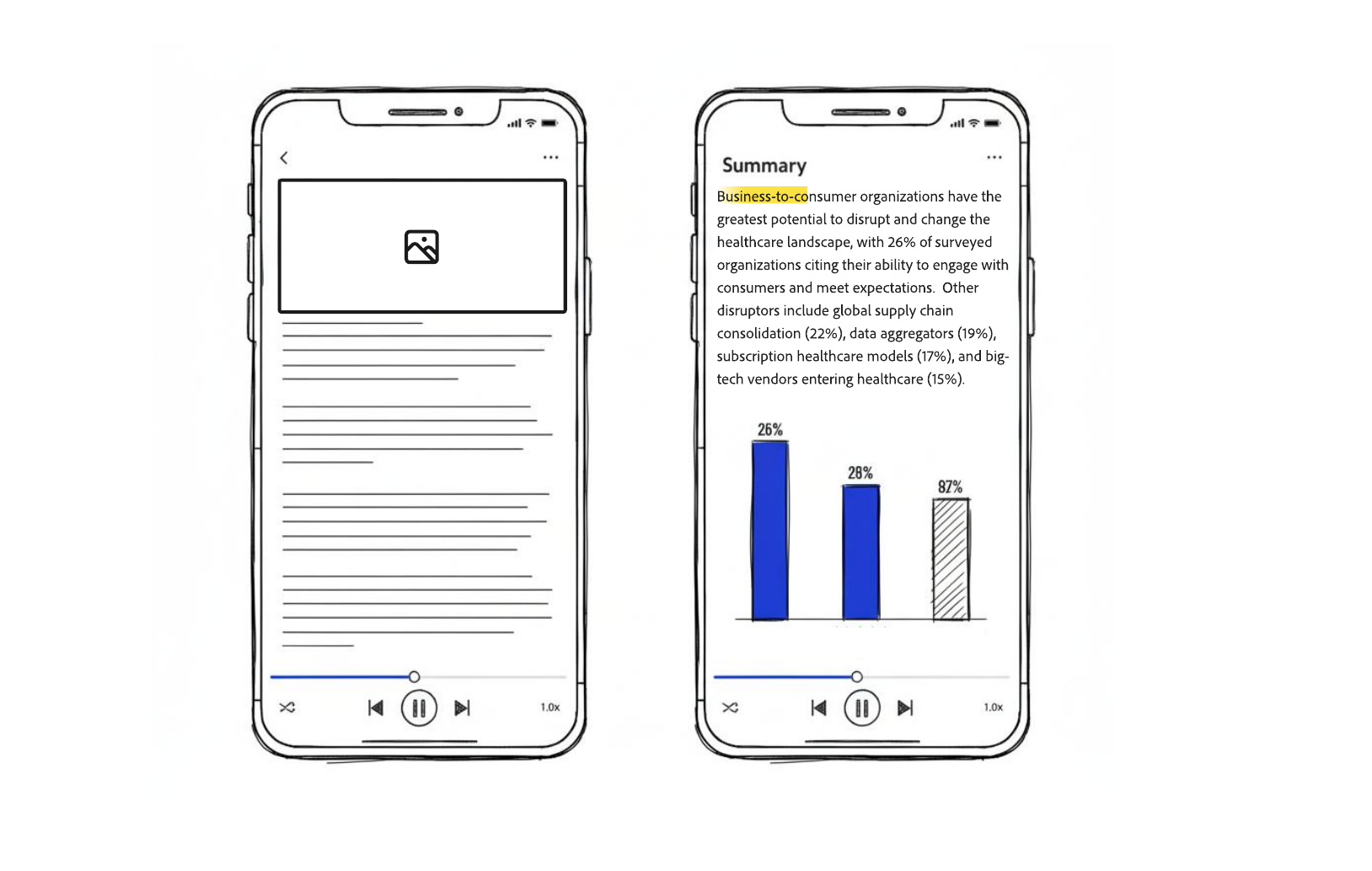}
    \vspace{-30pt}
    \caption{Concept C: Audio Assistant.}
    \vspace{20pt}
    \label{fig:concepts3}
\end{figure*}

\begin{figure*}[h!]
    \centering
    \includegraphics[width=0.8\linewidth]{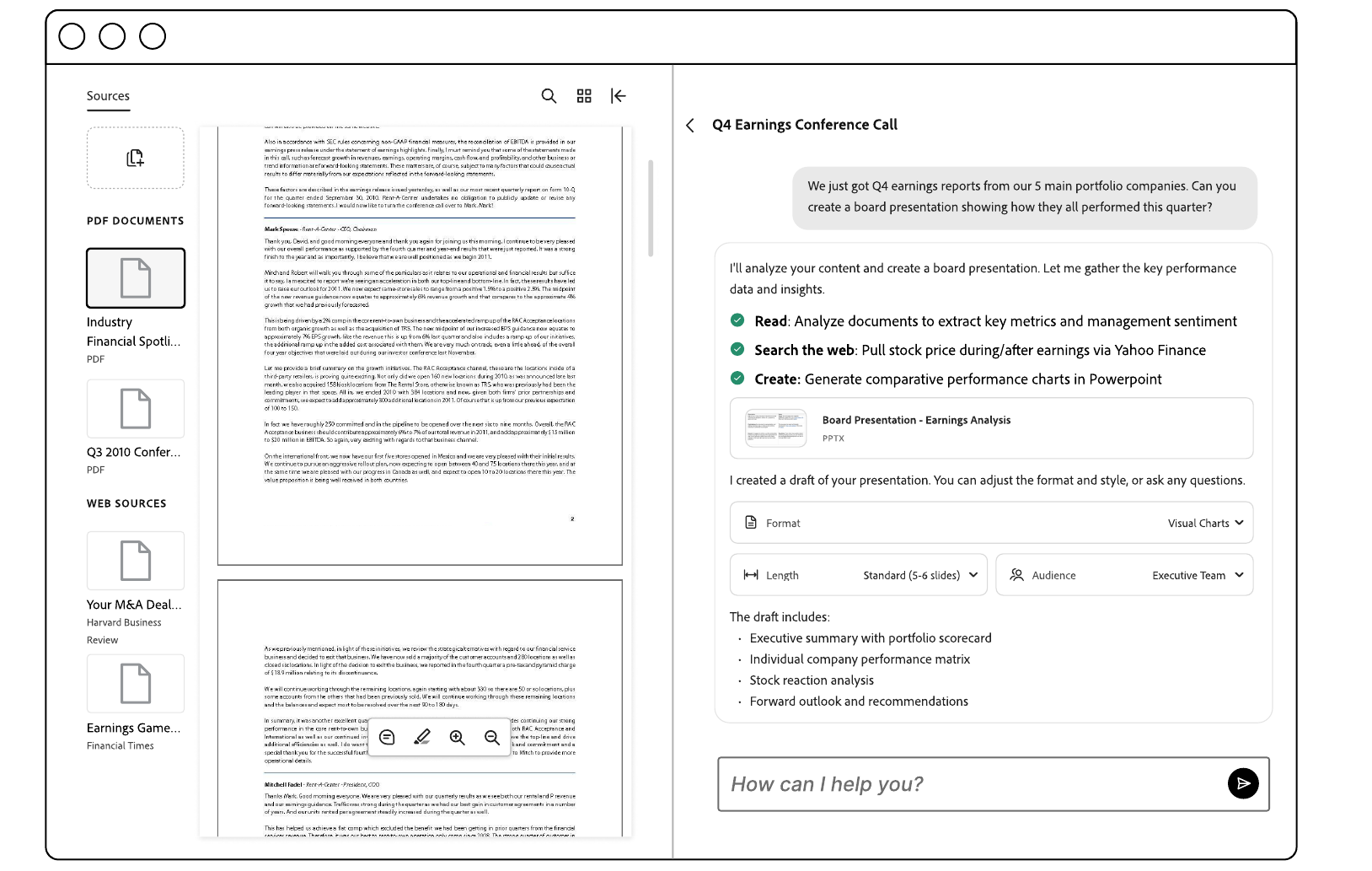}
    \vspace{-10pt}
    \caption{Concept D: Workflow Actions Assistant.}
    \label{fig:concepts4}
\end{figure*}

\added[id=jw1]{We include enlarged versions of figures~\ref{fig:concepts} in Figures~\ref{fig:concepts1} through~\ref{fig:concepts4}.}

\begin{table*}[ht!]
\centering
\begin{tabular}{lrrrr}
\toprule
\added[id=jw1]{\textbf{Comparison}} &
\added[id=jw1]{\textbf{Statistic}} &
\added[id=jw1]{\textbf{p-value}} &
\added[id=jw1]{\textbf{Lower CI}} &
\added[id=jw1]{\textbf{Upper CI}} \\
\midrule
\added[id=jw1]{(human -- interview-based)} & \added[id=jw1]{0.099} & \added[id=jw1]{0.093} & \added[id=jw1]{-0.011} & \added[id=jw1]{0.210} \\
\added[id=jw1]{(human -- scratchpad only)} & \added[id=jw1]{0.115} & \added[id=jw1]{0.037} & \added[id=jw1]{0.005}  & \added[id=jw1]{0.225} \\
\added[id=jw1]{(human -- no info)} & \added[id=jw1]{0.077} & \added[id=jw1]{0.275} & \added[id=jw1]{-0.034} & \added[id=jw1]{0.187} \\
\added[id=jw1]{(interview-based -- scratchpad only)} & \added[id=jw1]{0.015} & \added[id=jw1]{0.983} & \added[id=jw1]{-0.095} & \added[id=jw1]{0.126} \\
\added[id=jw1]{(interview-based -- no info)} & \added[id=jw1]{-0.023} & \added[id=jw1]{0.951} & \added[id=jw1]{-0.133} & \added[id=jw1]{0.087} \\
\added[id=jw1]{(scratchpad only -- no info)} & \added[id=jw1]{-0.038} & \added[id=jw1]{0.806} & \added[id=jw1]{-0.148} & \added[id=jw1]{0.072} \\
\bottomrule
\end{tabular}
\caption{\added[id=jw1]{Tukey's HSD Pairwise Group Comparisons (95.0\% Confidence Interval) for the Gwet's AC2 results in Figure~\ref{fig:main_mae_corr}.}}
\label{tab:tukey-hsd-2}
\end{table*}

\begin{figure*}
    \centering

    \includegraphics[width=0.65\linewidth]{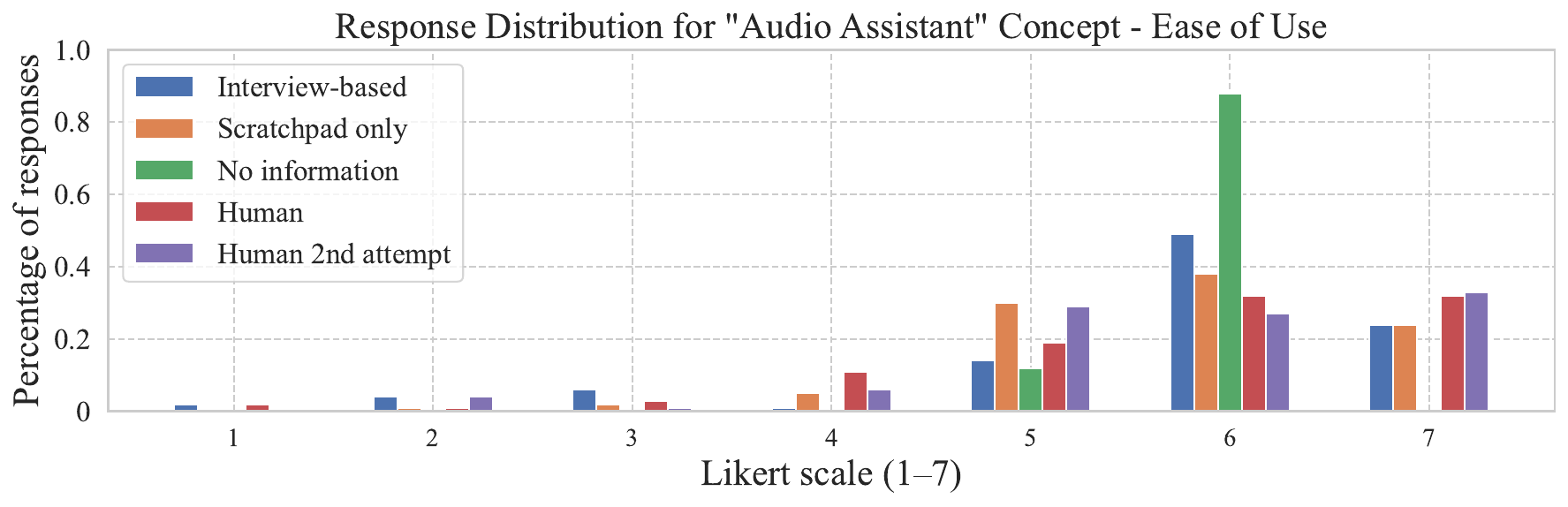}

    \vspace{1em}
    \includegraphics[width=0.65\linewidth]{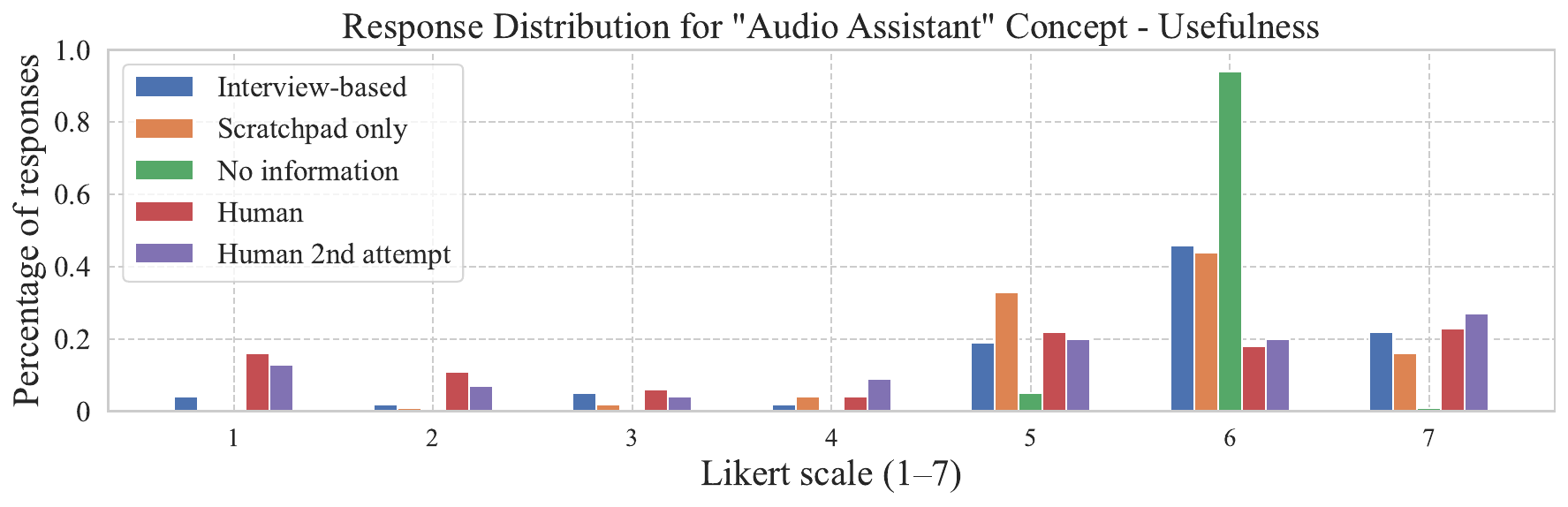}

    \vspace{1em}
    \includegraphics[width=0.65\linewidth]{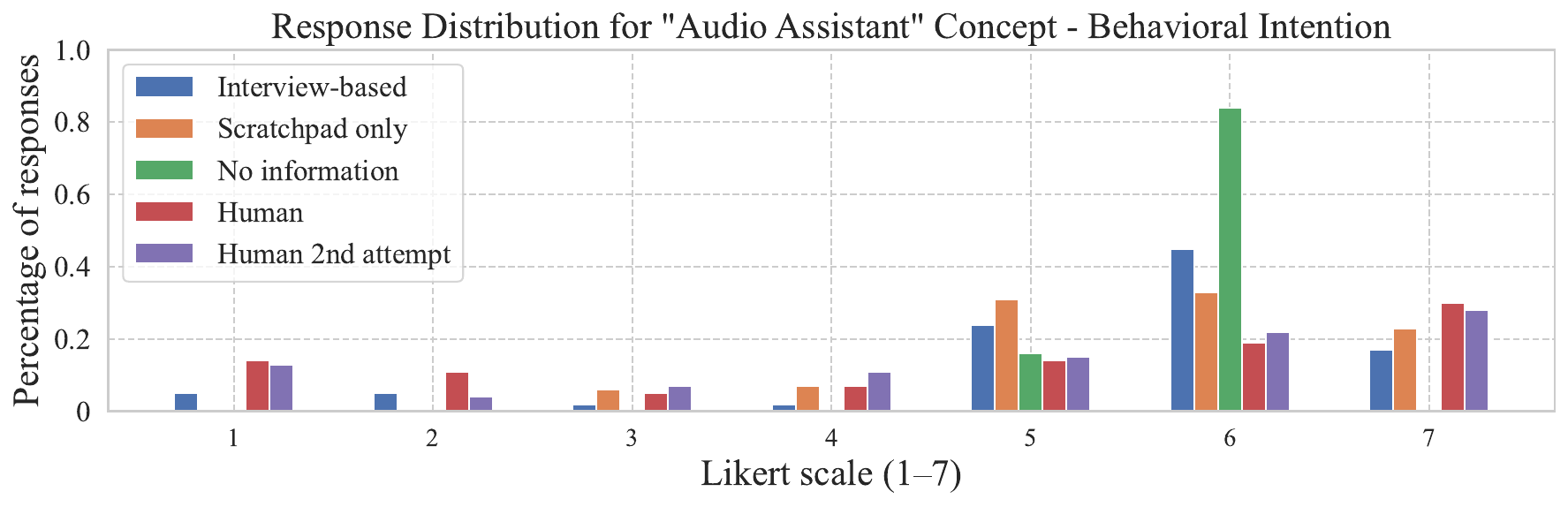}

    \vspace{1em}
    \includegraphics[width=0.65\linewidth]{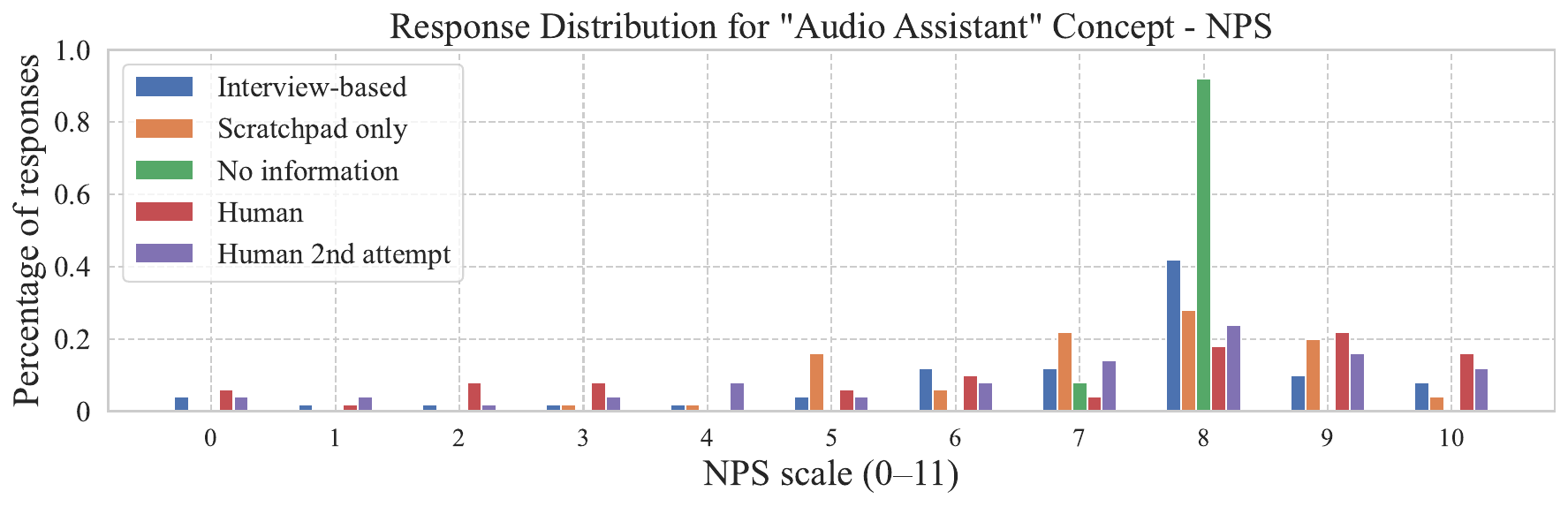}

    \vspace{1em}
    \resizebox{0.6\textwidth}{!}{%
\begin{tabular}{lcccc}
\toprule
& {Usefulness} 
& {Ease of Use} 
& {Behavioral Intention}
& {NPS} \\
\midrule
Human--Human 
& 0.330  
& 0.150 
& 0.200 
& 0.480 \\

Human--Interview-based 
& 0.820 
& 0.350 
& 0.710 
& 0.560 \\

Human--Scratchpad only 
& 1.030 
& 0.190
& 0.810 
& 0.780 \\

Human--No-information 
& 1.640 
& 0.810 
& 1.510 
& 1.980 \\
\bottomrule
\end{tabular}}

    \caption{Distribution of the four constructs for the ``Audio Assistant'' concept as well as the Wasserstein distance  comparing annotators themselves and the three different agent designs.}
    \label{fig:bigblock1}
\end{figure*}

\begin{figure*}
    \centering

    \includegraphics[width=0.65\linewidth]{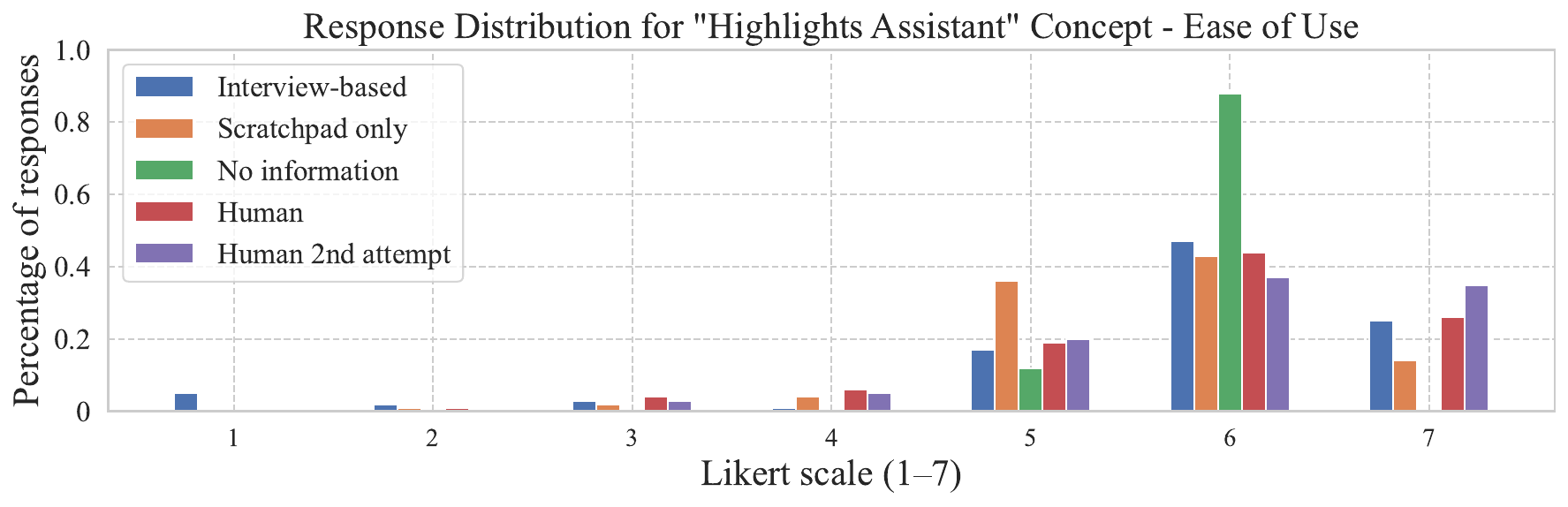}

    \vspace{1em}
    \includegraphics[width=0.65\linewidth]{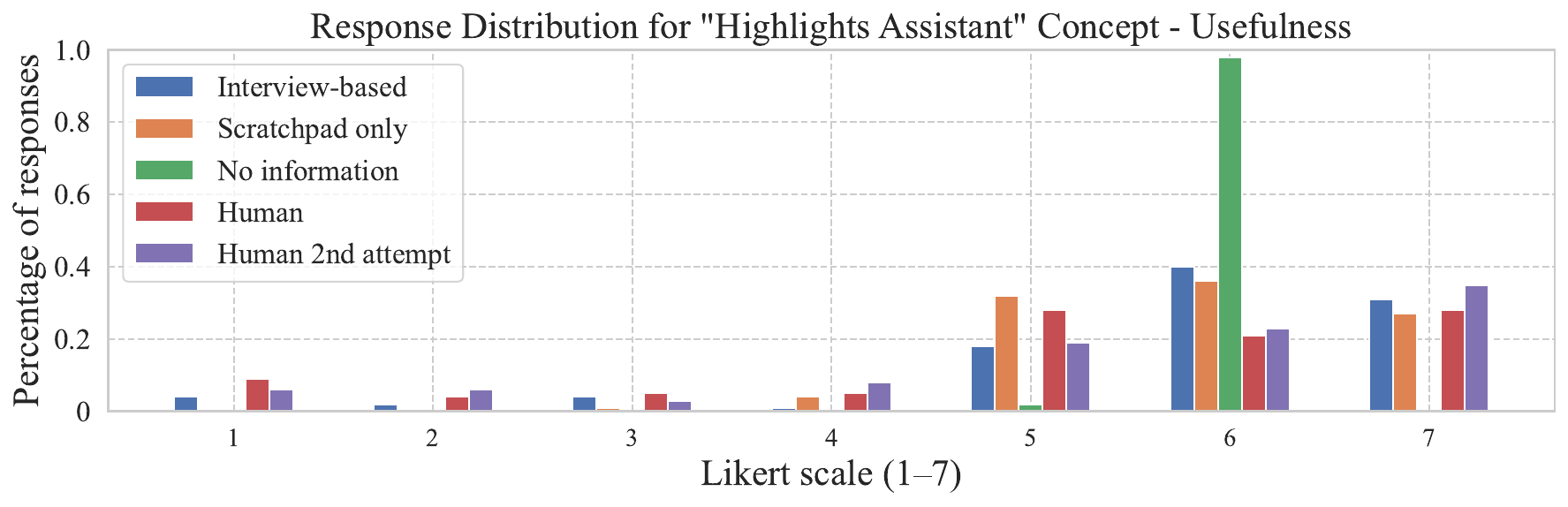}

    \vspace{1em}
    \includegraphics[width=0.65\linewidth]{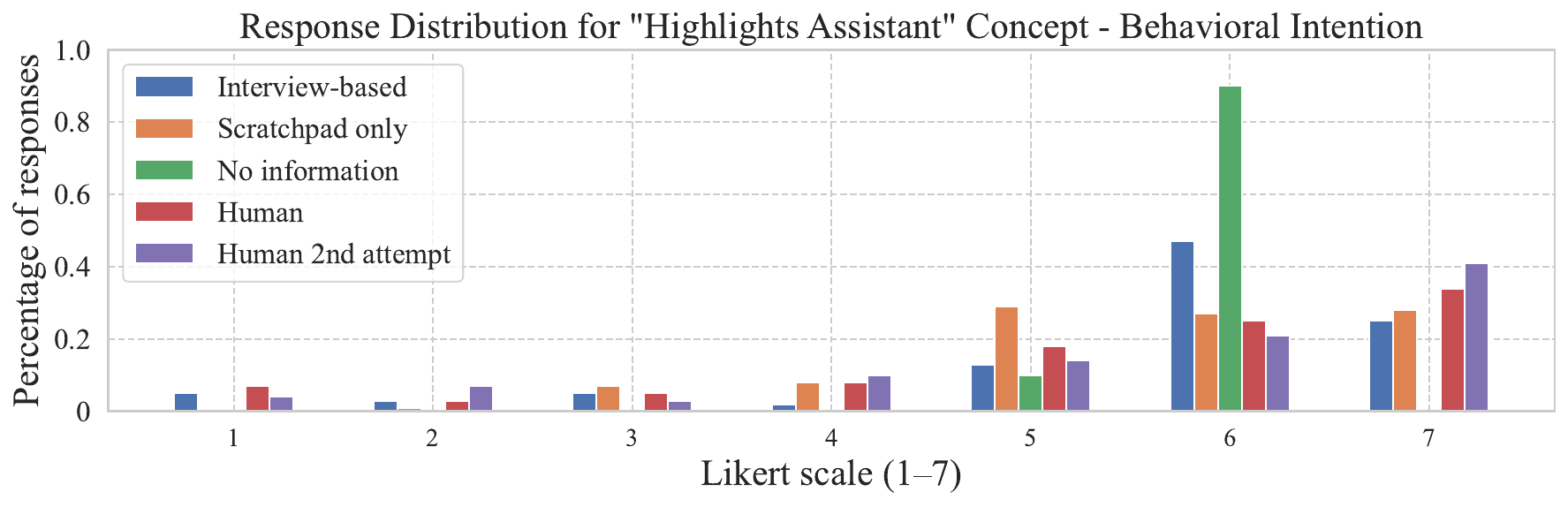}

    \vspace{1em}
    \includegraphics[width=0.65\linewidth]{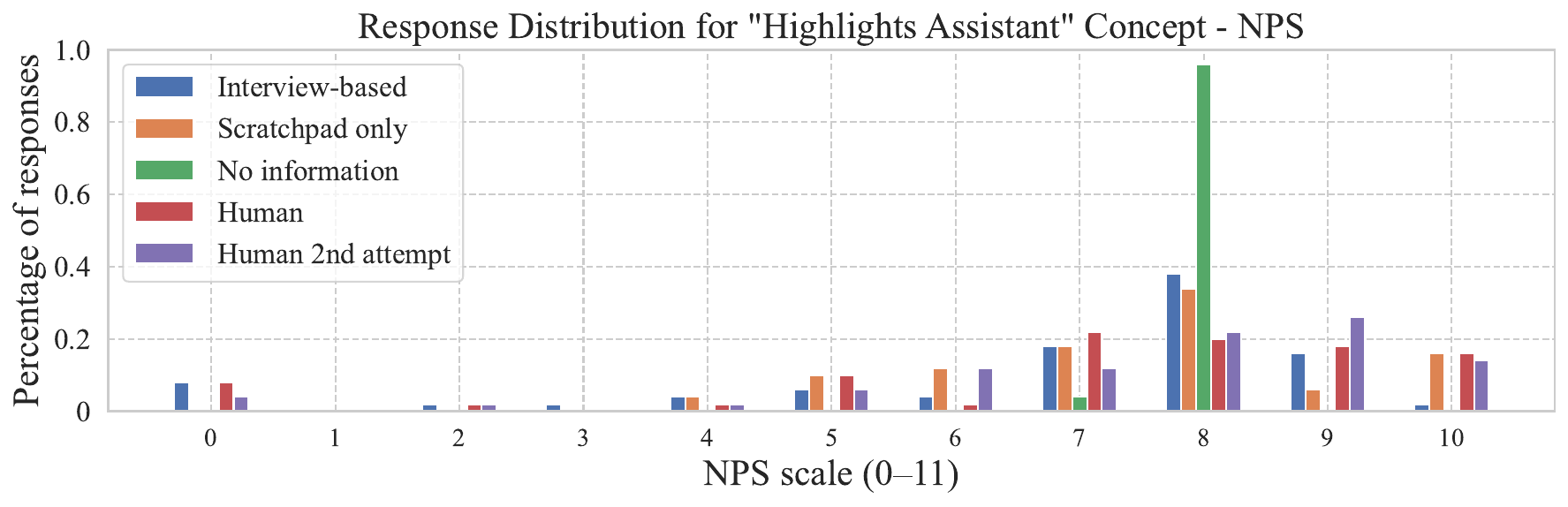}

    \vspace{1em}
    \resizebox{0.6\textwidth}{!}{%
\begin{tabular}{lcccc}
\toprule
& {Usefulness} 
& {Ease of Use} 
& {Behavioral Intention}
& {NPS} \\
\midrule
Human--Human 
& 0.230 
& 0.170
& 0.160
& 0.460 \\

Human--Interview-based 
& 0.420 
& 0.320 
& 0.400 
& 0.740 \\

Human--Scratchpad only 
& 0.630 
& 0.380
& 0.480 
& 0.500 \\

Human--No-information 
& 1.310 
& 0.620
& 1.220 
& 1.560 \\
\bottomrule
\end{tabular}}

    \caption{Distribution of the four constructs for the ``Highlight Assistant'' concept as well as the Wasserstein distance  comparing annotators themselves and the three different agent designs.}
    \label{fig:bigblock2}
\end{figure*}

\begin{figure*}
    \centering

    \includegraphics[width=0.65\linewidth]{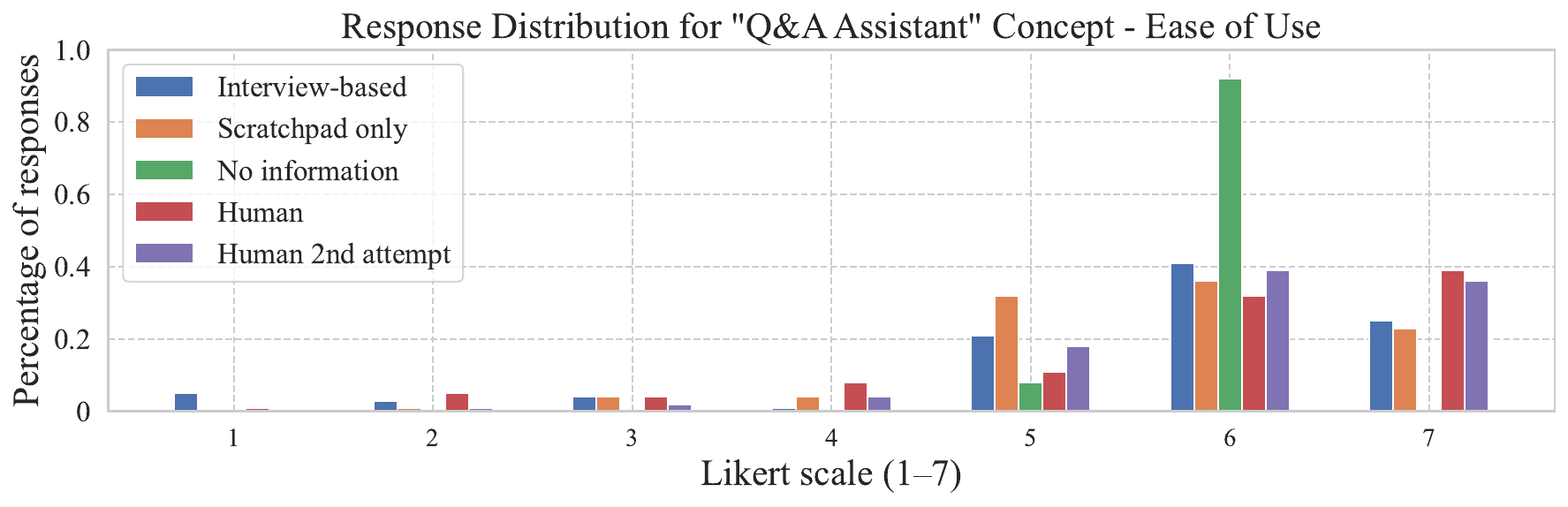}

    \vspace{1em}
    \includegraphics[width=0.65\linewidth]{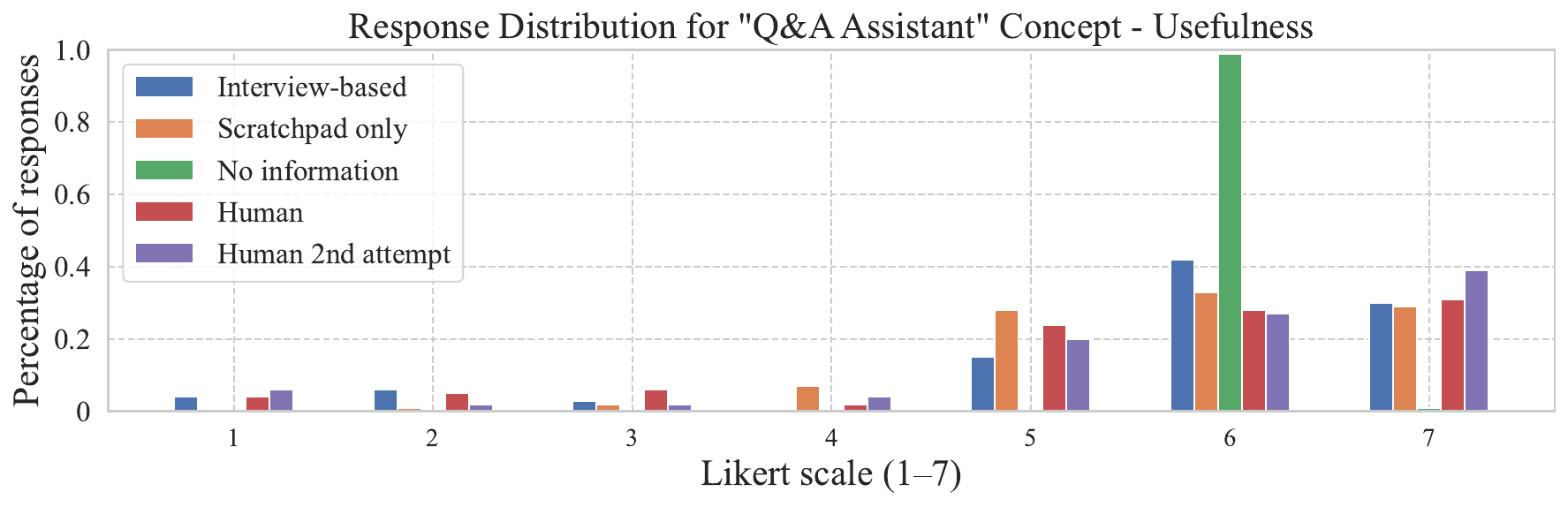}

    \vspace{1em}
    \includegraphics[width=0.65\linewidth]{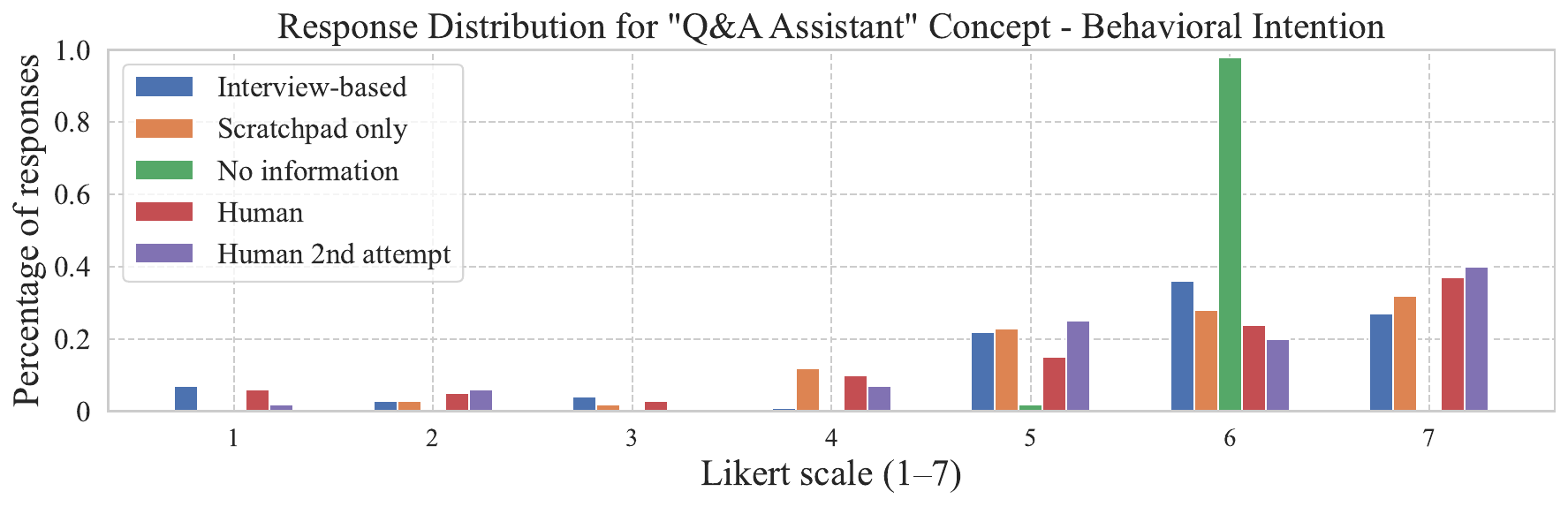}

    \vspace{1em}
    \includegraphics[width=0.65\linewidth]{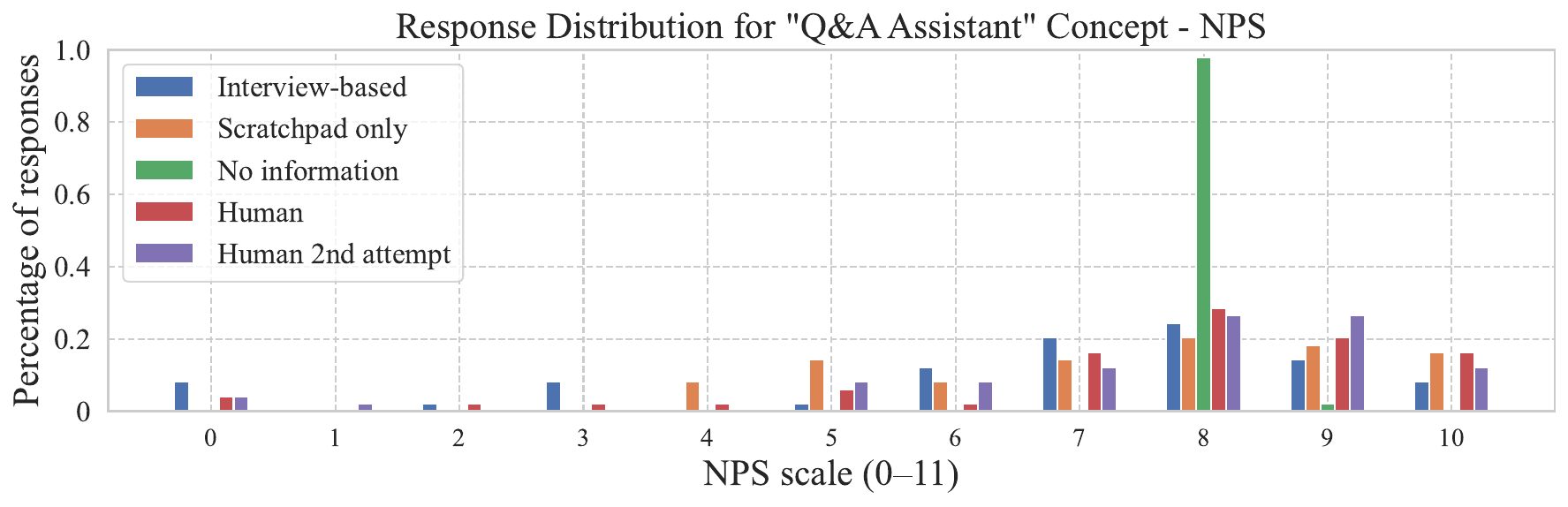}

    \vspace{1em}
    \resizebox{0.6\textwidth}{!}{%
\begin{tabular}{lcccc}
\toprule
& {Usefulness} 
& {Ease of Use} 
& {Behavioral Intention}
& {NPS} \\
\midrule
Human--Human 
& 0.260
& 0.310
& 0.260
& 0.204 \\

Human--Interview-based 
& 0.230 
& 0.470 
& 0.290 
& 0.898 \\

Human--Scratchpad only 
& 0.380 
& 0.330 
& 0.200 
& 0.592  \\

Human--No-information 
& 1.100 
& 0.640 
& 1.110 
& 1.490 \\
\bottomrule
\end{tabular}}

    \caption{Distribution of the four constructs for the ``Q\&A Assistant'' concept as well as the Wasserstein distance  comparing annotators themselves and the three different agent designs.}
    \label{fig:bigblock3}
\end{figure*}

\begin{figure*}
    \centering

    \includegraphics[width=0.65\linewidth]{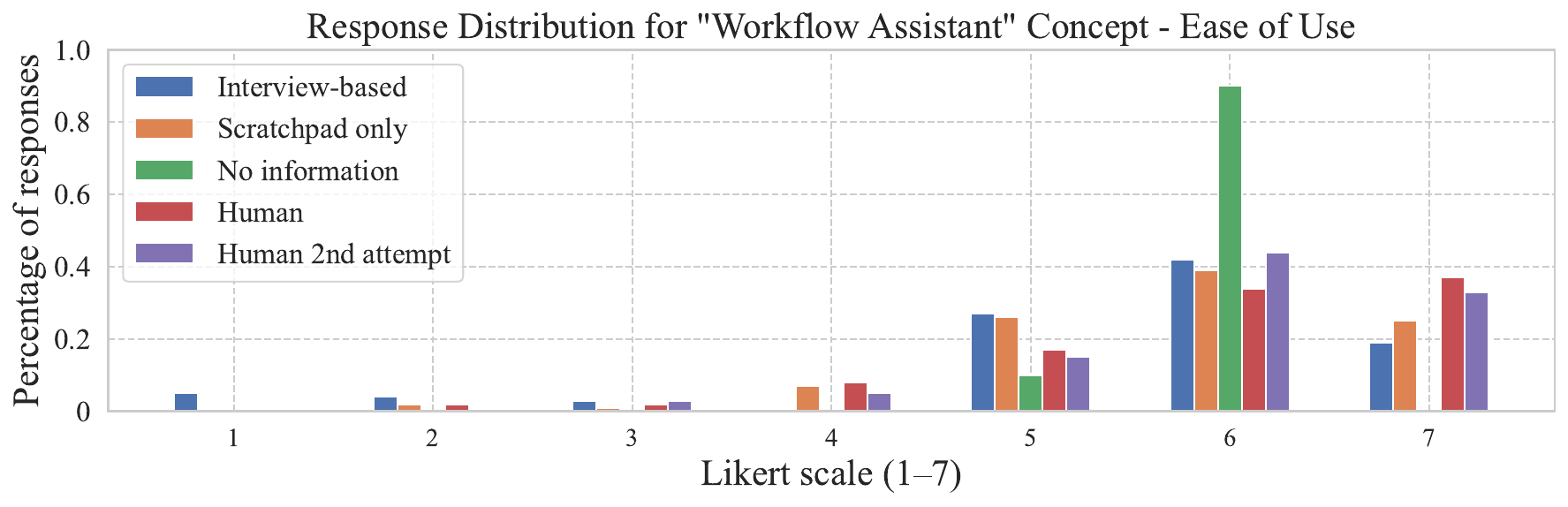}

    \vspace{1em}
    \includegraphics[width=0.65\linewidth]{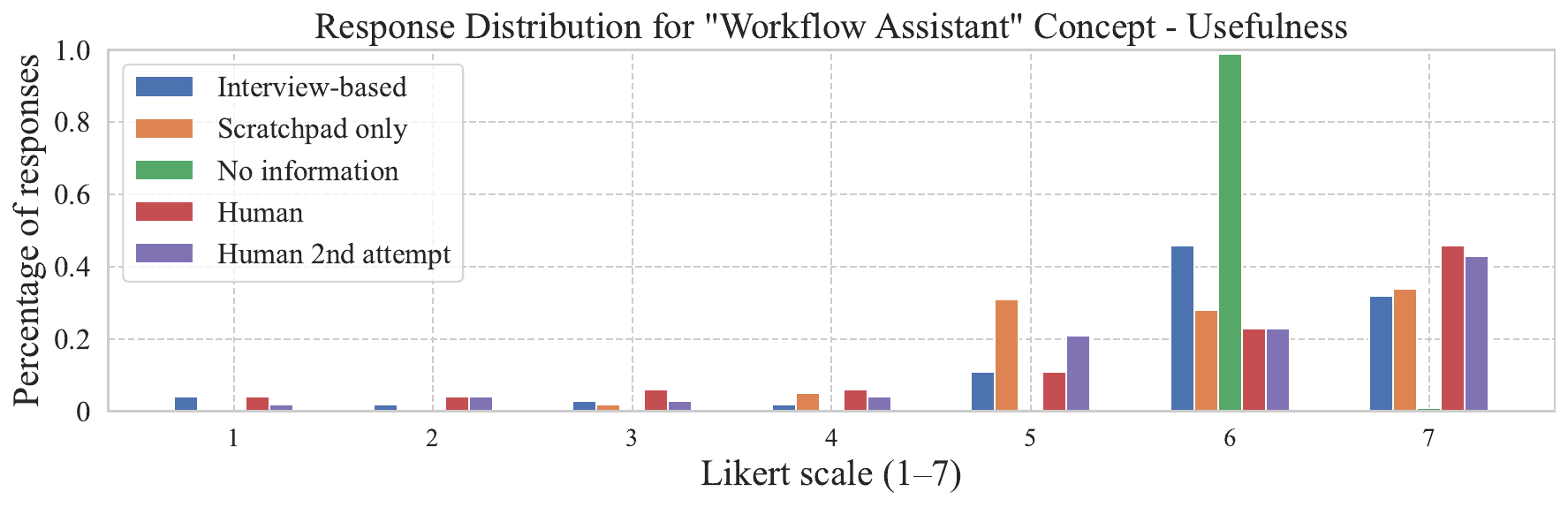}

    \vspace{1em}
    \includegraphics[width=0.65\linewidth]{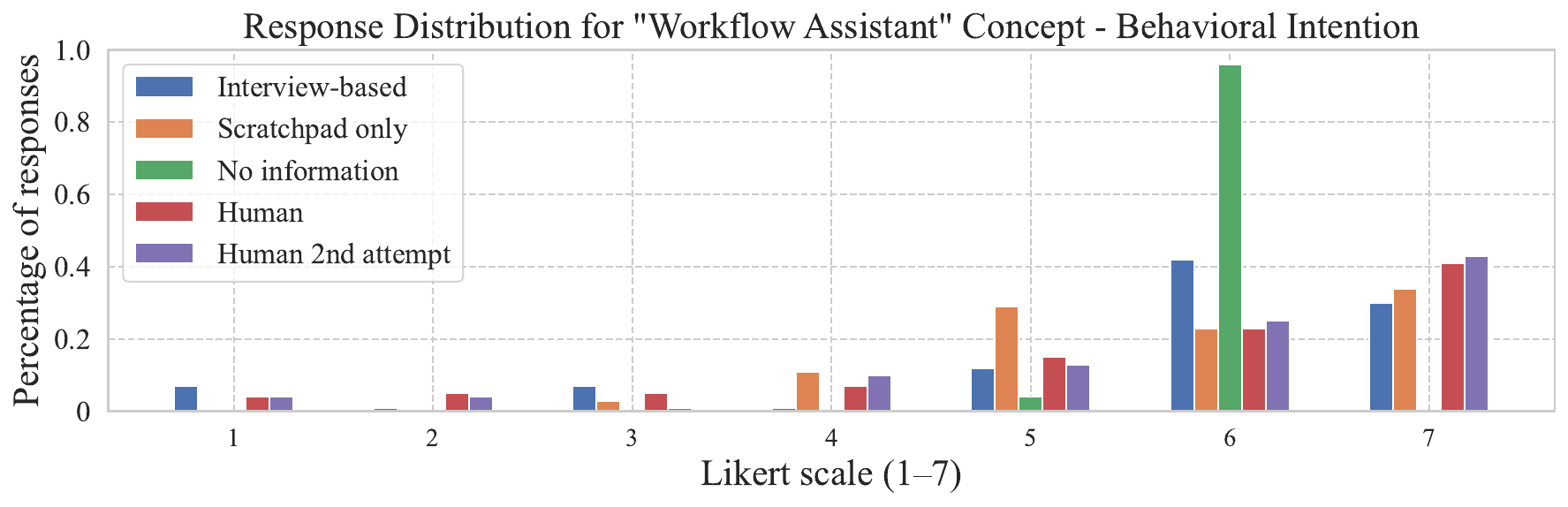}

    \vspace{1em}
    \includegraphics[width=0.65\linewidth]{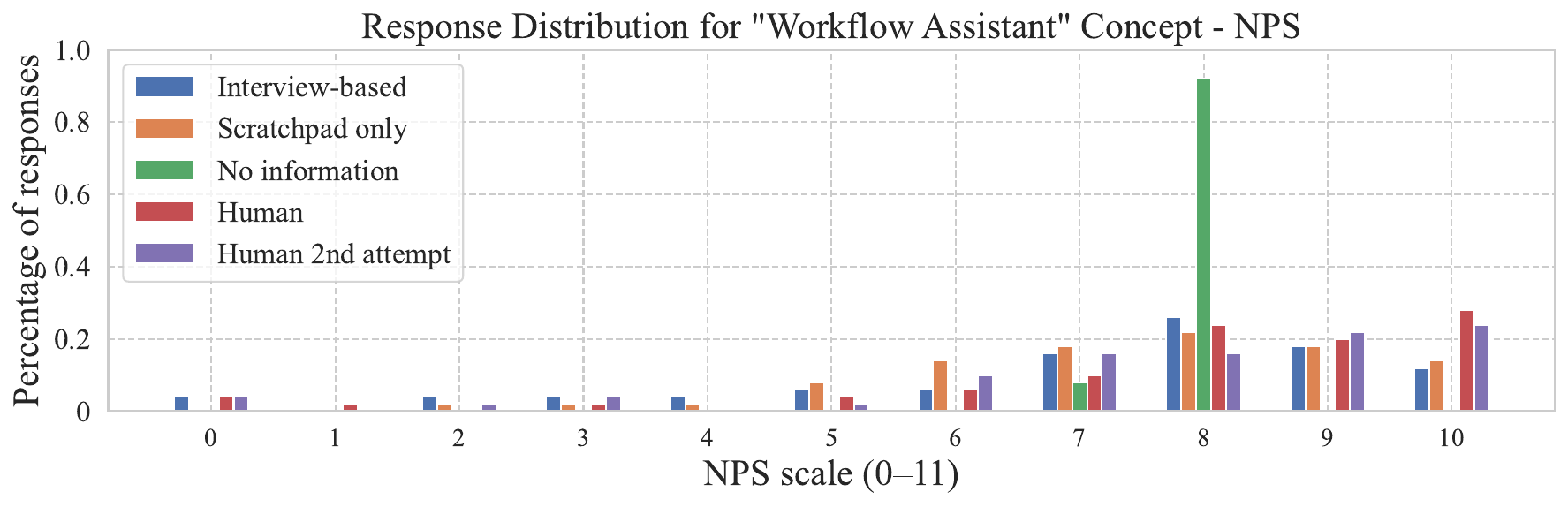}

    \vspace{1em}
    \resizebox{0.6\textwidth}{!}{%
\begin{tabular}{lcccc}
\toprule
& {Usefulness} 
& {Ease of Use} 
& {Behavioral Intention}
& {NPS} \\
\midrule
Human--Human 
& 0.220 
& 0.170 
& 0.140 
& 0.260  \\

Human--Interview-based 
& 0.270
& 0.570 
& 0.290 
& 0.600 \\

Human--Scratchpad only 
& 0.340
& 0.250 
& 0.430 
& 0.620  \\

Human--No-information 
& 1.060 
& 0.570 
& 1.110
& 1.680  \\
\bottomrule
\end{tabular}}

    \caption{Distribution of the four constructs for the ``Workflow Assistant'' concept as well as the Wasserstein distance  comparing annotators themselves and the three different agent designs.}
    \label{fig:bigblock4}
\end{figure*}

\end{document}